\documentclass[useAMS,usenatbib]{mn2e}
\usepackage[dvips]{graphicx}

\begin{document}

\title[Long Secondary Periods in Variable Red Giants]{Long Secondary
  Periods in Variable Red Giants}
\author[C. P. Nicholls et al.]{C. P. Nicholls$^{1}$\thanks{E-mail:
nicholls@mso.anu.edu.au (CPN); wood@mso.anu.edu.au (PRW); m.cioni@herts.ac.uk
  (M-RLC); soszynsk@.astrouw.edu.pl (IS)}, P. R. Wood$^{1}$\footnotemark[1],
  M.-R. L. Cioni$^{2}$\footnotemark[1] and I. Soszy\'nski$^{3}$\footnotemark[1]\\
$^{1}$Research School of Astronomy and Astrophysics, Australian National
University, Cotter Road, Weston Creek ACT 2611, Australia\\
$^{2}$Centre for Astrophysics Research, University of Hertfordshire, College Lane, Hatfield, AL10 9AB, UK\\
$^{3}$Warsaw University Observatory, Aleje Ujazdowskie 4, 00-478,
  Warsaw, Poland}

\date{Accepted 2009 July 15. Received 2009 June 23; in original form
  2009 April 22}

\pagerange{\pageref{firstpage}--\pageref{lastpage}} \pubyear{2009}

\maketitle

\label{firstpage}

\begin{abstract}

We present a study of a sample of LMC red giants exhibiting Long Secondary Periods (LSPs). We use radial velocities obtained from VLT spectral observations and MACHO and OGLE light curves to examine properties of the stars and to evaluate models for the cause of LSPs. This sample is much larger than
the combined previous studies of \cite{hinkle02} and \nocite{sequenceDstars} Wood, Olivier \& Kawaler (2004).

Binary and pulsation models have enjoyed much support in recent years. Assuming stellar pulsation, we calculate from the velocity curves that the typical fractional radius change over an LSP cycle is greater than 30 per cent. This should lead to large changes in $T_{\rm{eff}}$ that are not observed. Also, the small light amplitude of these stars seems inconsistent with the radius amplitude. We conclude that pulsation is not a likely explanation for the LSPs. The main alternative, physical movement of the star -- binary motion -- also has severe problems. If the velocity variations are due to binary motion, the distribution of the angle of periastron in our large sample of stars has a probability of $1.4 \times 10^{-3}$ that it comes from randomly aligned binary orbits. In addition, we calculate a typical companion mass of $0.09 \rm{M_{\odot}}$. Less than 1 per cent of low mass main sequence stars have companions near this mass (0.06 to 0.12 $\rm{M_{\odot}}$) whereas $\sim25$ to 50 per cent of low mass red giants end up with LSPs. We are unable to find a suitable model for the LSPs and conclude by listing their known properties.

\end{abstract}


\section{Introduction}

A subset of Long-Period Variable stars (LPVs) were found several decades ago to show a secondary period of variation, in addition to their 
primary pulsation \citep[e.g.][]{houk, payne-gaposchkin}. These Long Secondary Periods, or LSPs, exceed the primary period in length
by approximately one order of magnitude, a fact more recently confirmed by \nocite{wood99mn} Wood et al.\ (1999). Wood et al.\ showed that LPVs
fall on four distinct period-luminosity sequences A--D, and they found an additional sequence, E, of red giant binaries. These multiple sequences have been confirmed in subsequent studies, and a splitting of sequence B into two sequences has since been discovered \nocite{ogle04} \citep[Soszy\'nski et al.\ 2004a;][]{ita04, fraser05, oglep-l}. Long Secondary Periods occupy sequence D, the sequence corresponding to variations with the longest period. The primary pulsation of these stars is usually found on sequence B \nocite{wood99mn} (Wood et al.\ 1999). 

Approximately 25--50 per cent of Long-Period Variables show an LSP \nocite{wood99mn} \nocite{ogleellipsoidalmn} \citep[Wood et al.\ 1999; Soszy\'nski et al.\ 2004b;][]{percy, oglep-l, fraser08}. At present, there is no accepted explanation for LSPs. \nocite{wood99mn} Wood et al.\ (1999) initially proposed a model wherein the sequence D stars are semi-detached red giant binaries, as the length of their periods are consistent with those expected for binary systems with solar-mass components. Several other models have been proposed to explain LSPs, including radial and nonradial pulsation, and dust effects. However, \cite{sequenceDstars} demonstrate that there are problems with all of these models. It is clear that further investigation is required in order to discover the cause of LSPs. 

Here, we study a sample of sequence D stars in the Large Magellanic Cloud (LMC) in order to provide new constraints on models for the sequence D
phenomenon. In particular, we present new spectra taken with the ESO's Very Large Telescope (VLT). We derive velocity
curves, radii and effective temperatures from these spectra and compare them in phase with light curves and associated quantities from the MACHO and
OGLE databases.

Our study follows the methods of \cite{hinkle02} and \cite{sequenceDstars}, who also compared both light and radial velocity data for sequence D stars. However we use a larger dataset and have the advantage of high quality and simultaneous light and velocity data, which we hope will give a more complete and accurate picture of the behaviour of LSP variables. Since our sample comes from the LMC, we also have the advantage that the distances, and hence derived luminosities, are well-determined.

\section{Observations and Data Reduction}

We chose a region of high star density in the middle of the LMC bar for this study, in order to get a large sample of sequence D stars in the 20
arcmin field of the FLAMES multifibre system on the VLT. The field centre is at $05^h 28' 15'' -69^{\circ} 45' 43''$ J2000.
It contained a sample of 58 variable red giants exhibiting LSPs which were obtained from the MACHO Project database. It was important to
have simultaneous light and velocity data. Fortunately, OGLE light curves were taken at the times of our VLT observations. We have used both OGLE II
and OGLE III I-band light curves in this study. 

Spectra were obtained on 21 nights from 2003 November to 2006 March. The FLAMES/ GIRAFFE spectrograph (Pasquini et al.\ 2002) \nocite{pasquinimn}
with a grating setting of HR16 was used, giving spectra with a resolution of 23900 and a spectral interval of 693.7--725.0 nm. This region includes
the TiO bandhead at 705 nm, which can be used for spectral typing and to examine variation of $T_{\rm{eff}}$, as well as derivation of radial
velocities. An exposure time of 20 minutes was used for all spectra. The spectra were obtained in service mode. Unfortunately no observations were taken between May and August when the Large Magellanic Cloud \mbox{(RA = $05^{h}23'$)} was only observable near twilight, leaving a long Winter gap in 2004 and 2005. 

\begin{figure}
\begin{center}
\includegraphics[width=0.5\textwidth]{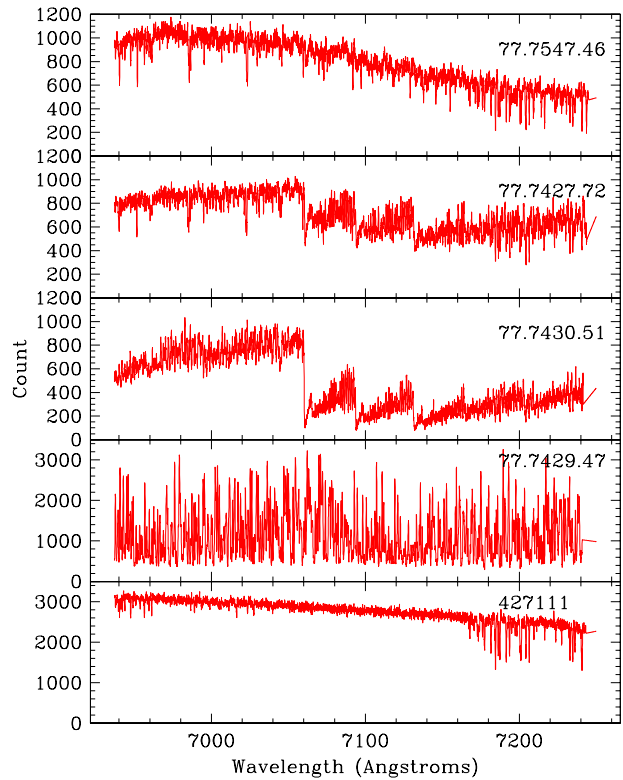}
\end{center}
\caption[Example Spectra]{Sample spectra from the VLT\@. \emph{Top panel}:
  The spectrum of an O-rich program star with no TiO bands. \emph{Second
  panel}: A program star with weak TiO bands. \emph{Third panel}: A
  program star with strong TiO bands. \emph{Fourth panel}: A C-rich
  program star. \emph{Bottom panel}: A telluric star. Program stars are identified by their MACHO numbers. Note the different
  scale of the y-axis in the lower two panels.}
\label{spectra}
\end{figure}

The raw spectra of the 58 sequence D stars, most of which have 21 observations from different dates were reduced using the FLAMES/GIRAFFE pipeline, or the Swiss reduction pipeline. A number of sample spectra are shown in Fig.~\ref{spectra}, where the TiO bandhead is clearly visible in the second and third panels. The majority of the program stars have spectra similar to those shown in the top three panels, i.e., they are oxygen-rich giants of spectral types M and K\@. However, the sample does contain a number of carbon-rich stars (C-stars), as shown in the fourth panel of Fig.~\ref{spectra}.

Using the \textsc{iraf} software, each spectrum was visually examined and any obvious cosmic rays removed. There  was a small percentage
of bad data: in particular, spectra taken on two dates (2005 March 17 and 2005 September 10) had very few counts. This was due to the weather
and seeing conditions at the time, and the spectra taken on these dates were discarded, leaving 19 good observations for most stars (some stars
had fewer spectra due to the fibre allocation process in FLAMES). A small number of other spectra were discarded due to a low number of
counts. These spectra did not fall on particular dates or particular target stars, and were distributed randomly throughout the dataset. Presumably, the low counts were due to the fibres for these stars being misaligned. 

Relative radial velocities were obtained through cross-correlation with the \textsc{iraf} task \emph{fxcor}. From each star, a spectrum
with a high number of counts and narrow lines was selected, and this acted as the template spectrum for the star's cross-correlation. The
cross-correlation was performed in the wavelength region 6950--7160 \AA, as this region is relatively free of telluric lines, as can be seen in
the lower panel of Fig.~\ref{spectra}. Errors in the radial velocities were also taken from \emph{fxcor}. The mean velocity error of the sample is $0.42 \rm{km\,s^{-1}}$, and errors typically range between 0.1 and $0.6 \rm{km\,s^{-1}}$.

Because of uncertainty about the accuracy of the velocity calibration resulting from the pipeline reduction, the spectra were checked to see if
any zero-point correction to the velocity was needed, before we calculated their absolute velocities. This procedure is now described.

\subsection{Testing the Relative Telluric Radial Velocities}

To check the velocity calibration, program stars were cross-correlated with a B star whose spectrum contained only telluric lines. Cross-correlation of a program spectrum to a telluric spectrum produces the relative telluric velocity of the two spectra, which should be zero if there are no calibration errors. This cross-correlation was performed in the wavelength region 7160--7220 \AA, as it has a large concentration of telluric lines (see Fig.~\ref{spectra}). 

For most dates of observation, the relative telluric radial velocities were found to be not significantly different from zero, and so
no systematic correction was applied to the spectra for those nights. However observations from two nights, with Heliocentric Julian Dates
(HJD) of 2453377 and 2453643, consistently showed a positive velocity offset from zero, so a zero-point correction for the velocities on these
dates was included for all stars. The velocity corrections for these dates were $-2.061 \rm{km\,s^{-1}}$ and $-0.729 \rm{km\,s^{-1}}$ respectively.

\subsection{The Effect of Telluric Lines on Relative Radial Velocities}

The highest concentration of telluric lines was found in the wavelength region 7160--7220 \AA, but there were a small number in the
program object region (6950--7160 \AA, see Fig.~\ref{spectra}). There was a possibility that these lines would affect the calculated radial
velocities, so this was examined. The telluric lines in the program object region of several program stars were removed using
\textsc{iraf}'s \emph{telluric} command. Then the program spectra with telluric lines removed were cross-correlated against each other in the
same way the original program spectra were. It was found that removing the telluric lines in this region had no significant effect on the
velocities or errors, so telluric line removal was not performed.

\subsection{Calculating the Absolute Radial Velocity}

Following the zero-point corrections, the absolute radial velocities of the program stars were calculated. First, the template of each program
star was cross-correlated to a star of known radial velocity, giving the absolute radial velocity of each template. Then by adding a keyword
\emph{`vhelio'} (which contained the absolute radial velocity) to the image header of the template, we could cross-correlate the spectra for
each star with its template to obtain the absolute radial velocities. 

For the O-rich program stars, the radial velocity reference used was $\alpha$ Cet. Its spectrum was taken using the echelle spectrograph (resolution 70000) on the late 74-inch telescope at Mount Stromlo Observatory, Canberra, Australia. For the C-stars, the reference used was the C-star X Vel, for which a spectrum was also taken with the 74-inch. The radial velocity used for $\alpha$ Cet was $-25.8 \rm{km\,s^{-1}}$ (taken from the Astronomical
Almanac)\nocite{almanac}. The velocity for the spectrum of X Vel ($-5.4 \rm{km\,s^{-1}}$) was obtained by cross-correlation with
$\alpha$ Cet (even though the spectral types are different, there was a strong cross-correlation peak due to common metal lines). 

Radial velocities as a function of heliocentric Julian date (HJD) for a sample of the sequence D stars are shown in Table~\ref{velocities}. Stars are identified by their MACHO numbers. Dashes denote dates for which there is no spectrum. The full table can be viewed in the online version of this paper.

\begin{table*}
\centering
\begin{minipage}{250mm}
\caption{Heliocentric Radial Velocities of Sequence D Stars ($\rm{km\,s^{-1}}$)}
\label{velocities}
\begin{tabular}{@{}lrrrrrrrrr@{}}
\hline
HJD  & 77.7427.41   & 77.7427.72   & 77.7428.120  & 77.7428.65   & 77.7429.100  & 77.7429.47   & 77.7429.64   & 77.7430.46   & 77.7430.51  \\
\hline
2954.85 & 240.85 & 273.53 & 273.61 & 297.74 & 314.55 & 280.43 & 278.15 & 244.02 & 266.31\\
3005.86 & 239.86 & 274.41 & 272.44 & 299.37 & 313.72 & 280.73 & 277.68 & 245.33 & 265.78\\
3067.60 & 241.32 & 275.08 & 272.69 & 298.12 & 313.20 & 279.58 & 277.01 & 244.83 & 267.10\\
3091.56 & 239.64 & 271.16 & 271.76 & 298.36 & 314.18 & 279.08 & 276.49 & 244.71 & 266.76\\
3280.86 & 241.59 & 273.70 & 272.70 & 295.91 & 314.24 & 278.73 &     -  & 241.53 & 264.68\\
3324.76 & 239.26 & 275.30 & 273.26 & 294.60 & 314.51 & 278.14 &     -  & 240.90 & 263.61\\
3344.80 & 240.04 & 274.01 & 273.34 & 295.59 & 315.15 & 277.93 &     -  & 242.23 & 263.88\\
3376.58 & 239.38 & 274.15 & 272.60 & 295.73 & 314.17 & 277.91 &     -  & 243.88 & 263.67\\
3418.69 & 238.11 & 273.68 & 272.59 & 296.53 & 315.73 & 277.24 &     -  & 244.06 & 265.99\\
3471.51 & 237.08 & 271.80 & 272.23 & 298.19 & 314.98 & 278.01 &     -  & 243.95 & 266.41\\
3623.88 &     -  &     -  &     -  &     -  &     -  &     -  &     -  &     -  &     - \\
3642.83 & 240.15 & 271.86 & 273.68 & 297.61 & 313.64 & 281.31 &     -  & 242.22 & 267.86\\
3644.82 & 239.89 & 271.50 & 273.67 & 297.36 & 313.43 & 281.03 &     -  & 241.70 & 267.60\\
3663.86 & 240.49 & 271.74 & 273.28 & 296.62 & 312.03 & 281.83 &     -  & 241.78 & 268.17\\
3664.86 & 240.62 & 271.92 & 273.53 & 296.83 & 311.85 & 281.99 &     -  & 241.45 & 268.16\\
3685.67 & 240.85 & 273.02 & 272.99 & 297.64 & 313.59 & 282.93 &     -  & 240.64 & 267.92\\
3709.79 & 240.50 & 269.51 & 273.40 & 297.15 & 314.57 & 284.06 &     -  & 242.48 & 266.48\\
3741.60 & 242.04 & 272.47 & 272.40 & 296.61 & 313.84 & 282.54 &     -  & 243.07 & 265.69\\
3768.62 & 241.31 & 273.48 & 274.19 & 297.47 & 313.53 & 280.68 &     -  & 242.40 & 266.54\\
3819.50 & 242.35 & 274.62 & 273.97 & 296.13 & 313.88 & 282.12 &     -  & 245.04 & 266.56\\
\hline
\end{tabular}
\end{minipage}
\end{table*}

\subsection{Correcting for the Short-Period Variation}
\label{correcting}

Because sequence D stars show two distinct modes of variability, it is necessary to correct the light and velocity data for the behaviour of
the short (primary) period in order to accurately study the LSP. We need simultaneous velocity and light data for this purpose, so we use the OGLE $I$ light curve in these procedures, even though it is missing the Southern Winter season.

First the OGLE $I$ light curve was interpolated and boxcar-smoothed over a time interval equal to the (primary) pulsation period. Once this LSP-only light curve was made, it was subtracted from the original light curve, leaving only the variations due to pulsation, $di$, or the `pulsation light curve'. 

Similarly to obtaining the pulsation light curve, we obtained the pulsation velocity curve by making a binary fit to the velocity
data using the period of the LSP\@. The deviations of the velocity data, $dv$, from this fit form the pulsation velocity curve. We emphasise that although a binary fit was used for the purpose of obtaining a smooth fit, binarity may not be the physical mechanism behind the LSP\@. 

For radial pulsation the light and velocity variations are shifted in phase relative to one another, so it is necessary to find the phase
shift between the pulsation light and velocity curves in order to find the amplitude relation between the short-period light and velocity. 

The phase correction was found by plotting the pulsation light versus pulsation velocity. By varying the phase shift of this plot, and judging by eye at
which phase shift the pulsation light and velocity showed the clearest correlation, the best phase shift was found to be $\sim$0.25 of a cycle. This
means that most often the  $I$ light minimum occurs close to the same phase as the mean rising radial velocity. Lebzelter, Kiss \& Hinkle (2000)
\nocite{lebzelter00} and \cite{lebzelter} find that the phase shift between the light and velocity variations of pulsating SRVs is
0.5. This means maximum radial velocity occurs at minimum light, and vice versa. The phase shift is discussed further below.

Next, the slope of the light--velocity amplitude correlation for pulsation was measured, to give the velocity correction. From examination of
several stars with relatively large pulsation amplitudes, it was concluded that the velocity correction was $\sim$15 $\rm{km\,s^{-1}\,mag^{-1}}$ in I. Generally, the correction to the LSP velocity resulting from these processes is small, less than $0.6 \rm{km\,s^{-1}}$, as the median value of $di$ is $\sim$0.04 mag. 

In order to further investigate the phase relation between the velocity and light curves, we constructed a model semi-regular pulsator using the
nonlinear pulsation code described in \cite{kellerwood}. The model is a first overtone pulsator, appropriate for a sequence B variable. The model
oscillations should be similar to the primary oscillation in the sequence D variables studied here, and the semi-regular variables
studied by \cite{lebzelter}. The model had $M=1.5 \rm{M_{\odot}}$, $L=3000 \rm{L_{\odot}}$, $T_{\rm{eff}}=3740 K$, helium abundance $Y=0.3$ and metallicity $Z=0.004$. $M_{V}$ and $M_{I}$ were computed from the M star model atmospheres of \cite{houdashelt}, assuming $[Fe/H] = -0.5$.

\begin{figure}
\begin{center}
\includegraphics[width=0.5\textwidth]{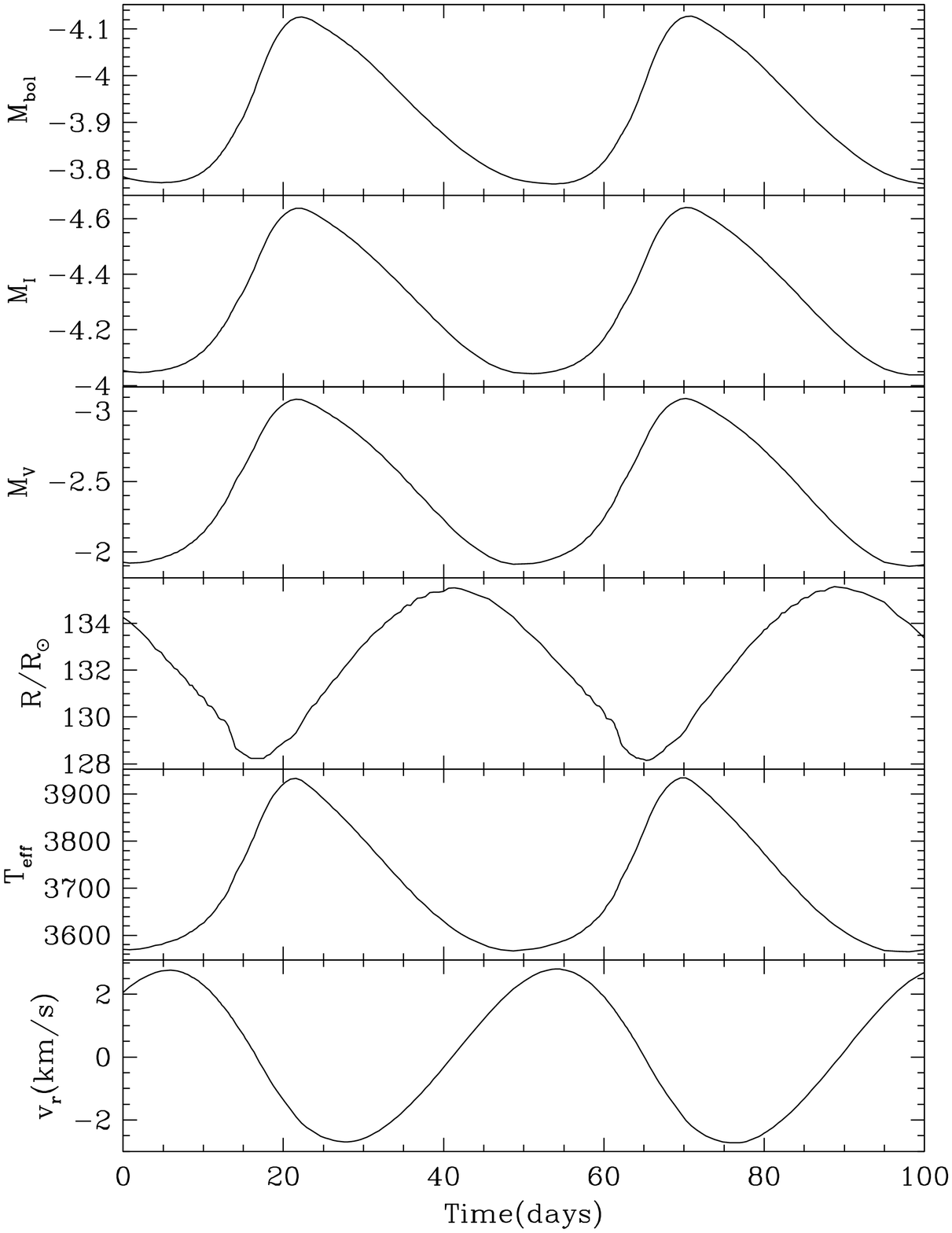}
\end{center}
\caption[Model red giant variable pulsating in the first overtone
  mode.]{$M_{bol}$, $M_{I}$, $M_{V}$, $R$/$R_{\odot}$, $T_{\rm{eff}}$ 
  and $v_r$ plotted against time in a model red giant variable pulsating
  in the first overtone mode. The radius $R$ is defined at optical depth
  $\frac{2}{3}$ and $v_r$ is the radial velocity seen by a distant
  observer. It is defined as $-v$/1.4, where $v$ is the pulsation
  velocity, relative to the centre of the star, of matter near optical
  depth $\frac{2}{3}$. The factor 1.4 is the correction factor from
  observed to pulsation velocities for red giant pulsators \citep{scholzwood}.}
\label{sr_model}
\end{figure}

The model light and radial velocity curves are shown in Fig.~\ref{sr_model}. The graph also shows the time variation of $T_{\rm{eff}}$ and radius for
comparison with observational estimates of these quantities in later sections.

The model shows that minimum light occurs between mean increasing radial velocity (as estimated here) and maximum radial velocity \citep[as estimated
by][]{lebzelter}. The model also predicts a ratio of velocity to $I$ amplitude of $9.2 \rm{km\,s^{-1}\,mag^{-1}}$, somewhat smaller than the estimate
of $15 \rm{km\,s^{-1}\,mag^{-1}}$ given here. Given the small size of the corrections to velocity that we have applied, the uncertainty in the
relative phase and amplitude ratio of velocity and $I$ amplitude will not significantly affect the resulting velocity curve of the sequence D variability.

\subsection{Obtaining LSP Parameters and Narrowing the Sample}
\label{obtaining}

\begin{figure*}
\begin{center}
\includegraphics[width=0.9\textwidth]{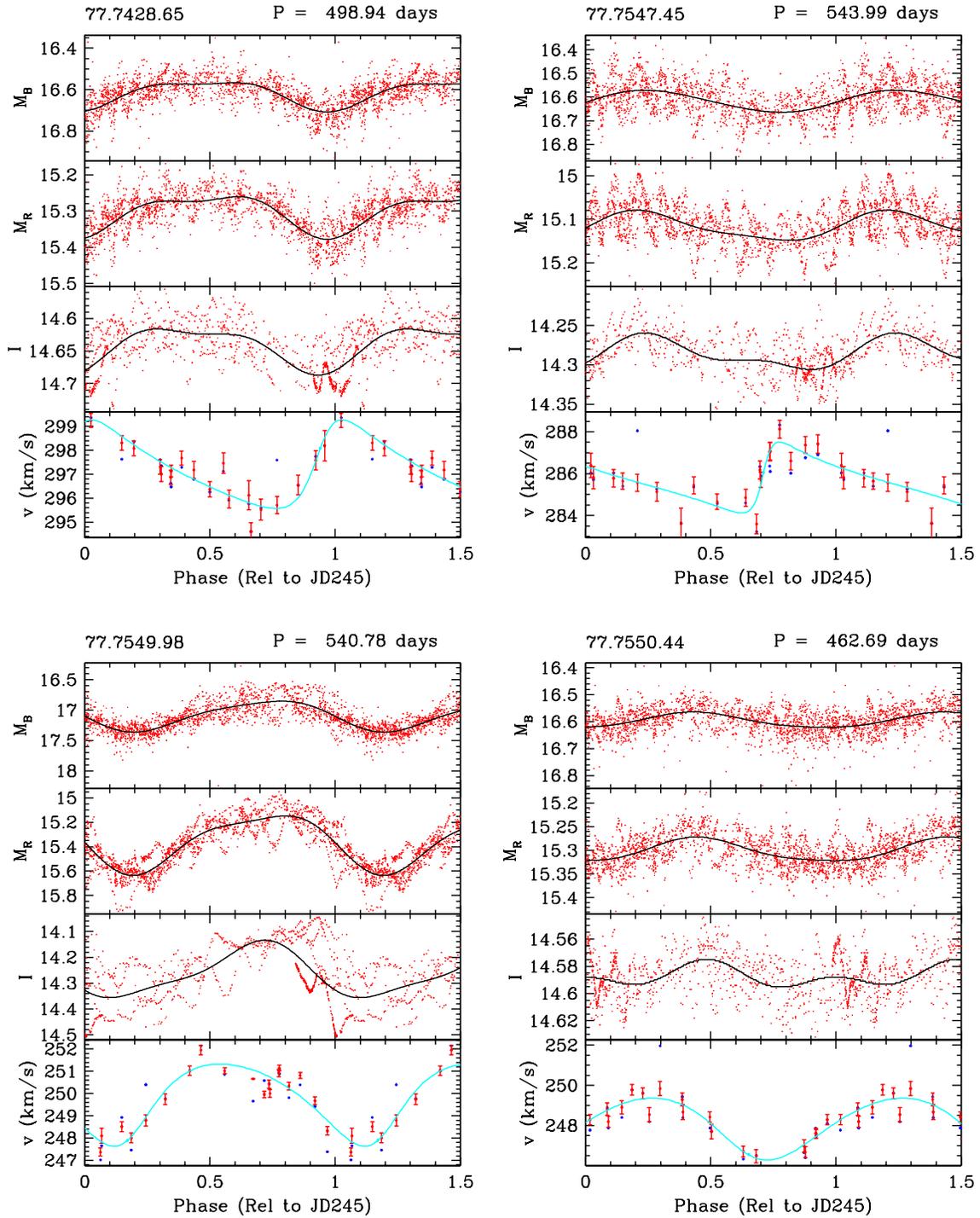}
\end{center}
\caption[Fourier Fits to the Light Data and a Binary Fit to the Velocity
  Data]{Light and velocity curves plotted against phase for four sequence D
  stars. The blue velocity points without error bars are the radial velocities without corrections and the red points with error bars are the radial velocities that have been corrected for pulsation and had specific date corrections. Also shown are the Fourier fits to the light data (black
  lines), and the Binary fits to the velocity data (cyan lines), both
  made by \textsc{Fitall}.} 
\label{fitting}
\end{figure*}

Following the corrections described above, we were now in a position to analyse the light and velocity variations associated with the Long Secondary Period. A binary fit was made to the corrected velocity data of the LSP using a Fortran program, \textsc{Fitall} to obtain the parameters of the velocity curve. \textsc{Fitall} also made a Fourier series fit (with a frequency f and one harmonic 2f) to the MACHO $M_{B}$, $M_{R}$ and OGLE $I$ light curves, and to ($M_{B}-M_{R}$). Some examples of the fits to the light and velocity data made by \textsc{Fitall} can be seen in Fig.~\ref{fitting}.

Due to the poor quality of the data for some stars, the original sample of 58 sequence D stars had to be reduced. Some stars had large velocity
errors due to poor-quality or few spectra or no OGLE light curve, so they were excluded from the sample. A few stars had LSPs near 365 days
leading to a large unfilled gap in the velocity curve, and could not be analysed. 

Next, for each of the remaining stars we judged by eye the quality of each star's binary fit, and eliminated all stars judged to have poor fits
to their velocity data. Finally, we were left with a reduced sample of 30 sequence D stars, from the original 58. The parameters of the velocity fit for these 30 stars are shown in Table~\ref{orbital}, where $\gamma$ is the system velocity, K is the velocity semiamplitude, e is the eccentricity, $\omega$ is the angle of periastron, T is the date of periastron, P is the period, and f(m) is the mass function, given by 
\begin{equation}
f(m) = \frac{K^3 P}{2 \pi G} = \frac{m^3 \sin ^3 i}{(m+M)^2}
\end{equation} 
in which $M$ is the mass of the star, and $m$ the mass of its companion.  

\begin{table*}
\centering
\caption{Orbital Elements}
\label{orbital}
\begin{tabular}{lrrrrrrrl}
\hline
\multicolumn{1}{c}{Star}  &  \multicolumn{1}{c}{$\gamma$} &  \multicolumn{1}{c}{K} &  \multicolumn{1}{c}{e}  &  \multicolumn{1}{c}{$\omega$} &  \multicolumn{1}{c}{T}  &  \multicolumn{1}{c}{P}  &  \multicolumn{1}{c}{f(m)} & \multicolumn{1}{l}{No.}\\
  &  \multicolumn{1}{c}{($\rm{km\,s^{-1}}$)}  &  \multicolumn{1}{c}{($\rm{km\,s^{-1}}$)}  &  &  \multicolumn{1}{c}{(deg)}  &  \multicolumn{1}{c}{(HJD)}  &  \multicolumn{1}{c}{(days)}  &  \multicolumn{1}{c}{($10^{-4}\rm{M_{\odot}}$)} & \multicolumn{1}{l}{obs.}\\
\hline
77.7427.41 & 239.812 $\pm$ 0.144 &  1.870 $\pm$ 0.175 & 0.436 $\pm$ 0.139 & 189.68 $\pm$  6.91 & 3491.19 $\pm$ 13.82 & 776.18 &    3.835 $\pm$    1.377 &  19 \\
77.7427.72 & 273.087 $\pm$ 0.186 &  1.626 $\pm$ 0.271 & 0.547 $\pm$ 0.182 & 257.71 $\pm$  6.59 & 3251.47 $\pm$ 10.27 & 488.67 &    1.277 $\pm$    0.840 &  19 \\
77.7428.65 & 297.159 $\pm$ 0.116 &  1.850 $\pm$ 0.095 & 0.402 $\pm$ 0.115 & 290.72 $\pm$  4.53 & 3462.77 $\pm$  4.97 & 498.94 &    2.512 $\pm$    0.567 &  19 \\
77.7429.100 & 313.955 $\pm$ 0.145 &  0.966 $\pm$ 0.188 & 0.187 $\pm$ 0.145 & 182.16 $\pm$ 12.49 & 3166.86 $\pm$ 16.46 & 473.24 &    0.419 $\pm$    0.248 &  19 \\
77.7429.47 & 280.326 $\pm$ 0.108 &  2.271 $\pm$ 0.120 & 0.475 $\pm$ 0.108 & 274.88 $\pm$  3.59 & 3547.93 $\pm$  9.50 & 1078.33 &    8.910 $\pm$    2.264 &  19 \\
77.7430.46 & 243.159 $\pm$ 0.129 &  1.687 $\pm$ 0.161 & 0.195 $\pm$ 0.127 & 279.99 $\pm$  5.57 & 3370.88 $\pm$  6.56 & 416.45 &    1.957 $\pm$    0.579 &  19 \\
77.7430.51 & 266.274 $\pm$ 0.212 &  2.203 $\pm$ 0.137 & 0.583 $\pm$ 0.202 & 227.13 $\pm$  3.50 & 4148.37 $\pm$ 11.12 & 798.34 &    4.737 $\pm$    2.686 &  18 \\
77.7547.44 & 271.204 $\pm$ 0.148 &  2.150 $\pm$ 0.163 & 0.151 $\pm$ 0.147 & 87.27 $\pm$  4.59 & 3314.19 $\pm$ 10.26 & 870.06 &    8.654 $\pm$    2.053 &  19 \\
77.7547.45 & 285.675 $\pm$ 0.142 &  1.697 $\pm$ 0.181 & 0.590 $\pm$ 0.141 & 278.00 $\pm$  3.93 & 3106.62 $\pm$  6.16 & 543.99 &    1.450 $\pm$    0.725 &  19 \\
77.7549.42 & 263.993 $\pm$ 0.146 &  1.758 $\pm$ 0.235 & 0.274 $\pm$ 0.144 & 225.45 $\pm$  3.01 & 3668.10 $\pm$  7.03 & 868.76 &    4.351 $\pm$    1.832 &  18 \\
77.7549.46 & 263.453 $\pm$ 0.188 &  1.542 $\pm$ 0.225 & 0.201 $\pm$ 0.186 & 242.52 $\pm$ 12.16 & 3169.49 $\pm$ 20.31 & 742.68 &    2.651 $\pm$    1.201 &  19 \\
77.7549.98 & 249.842 $\pm$ 0.123 &  1.848 $\pm$ 0.101 & 0.232 $\pm$ 0.123 & 210.85 $\pm$  5.25 & 3336.64 $\pm$  7.44 & 540.78 &    3.254 $\pm$    0.609 &  19 \\
77.7550.43 & 244.373 $\pm$ 0.170 &  1.743 $\pm$ 0.235 & 0.391 $\pm$ 0.169 & 255.86 $\pm$  7.39 & 3371.03 $\pm$  9.06 & 452.37 &    1.936 $\pm$    0.906 &  18 \\
77.7550.44 & 247.976 $\pm$ 0.085 &  1.536 $\pm$ 0.129 & 0.118 $\pm$ 0.085 & 146.30 $\pm$  4.21 & 3540.15 $\pm$  5.29 & 462.69 &    1.701 $\pm$    0.432 &  19 \\
77.7552.111 & 273.144 $\pm$ 0.157 &  1.800 $\pm$ 0.180 & 0.217 $\pm$ 0.156 & 169.75 $\pm$  6.69 & 3433.09 $\pm$  8.76 & 462.96 &    2.604 $\pm$    0.828 &  19 \\
77.7667.918 & 260.681 $\pm$ 0.171 &  2.846 $\pm$ 0.130 & 0.409 $\pm$ 0.166 & 192.60 $\pm$  5.54 & 3182.52 $\pm$  4.05 & 310.17 &    5.632 $\pm$    1.576 &  19 \\
77.7669.1027 & 268.918 $\pm$ 0.202 &  3.360 $\pm$ 0.174 & 0.351 $\pm$ 0.202 & 276.46 $\pm$  5.69 & 3167.47 $\pm$  3.76 & 266.77 &    8.608 $\pm$    2.477 &  19 \\
77.7669.1028 & 220.264 $\pm$ 0.125 &  1.574 $\pm$ 0.186 & 0.171 $\pm$ 0.125 & 182.39 $\pm$  5.47 & 3220.97 $\pm$  9.44 & 623.22 &    2.409 $\pm$    0.868 &  19 \\
77.7669.973 & 273.015 $\pm$ 0.131 &  2.378 $\pm$ 0.215 & 0.257 $\pm$ 0.125 & 328.59 $\pm$  2.37 & 3193.90 $\pm$  2.05 & 367.77 &    4.622 $\pm$    1.340 &  19 \\
77.7669.991 & 302.589 $\pm$ 0.094 &  1.271 $\pm$ 0.105 & 0.529 $\pm$ 0.090 & 191.96 $\pm$  7.42 & 3091.45 $\pm$  4.05 & 241.77 &    0.315 $\pm$    0.100 &  19 \\
77.7671.282 & 239.954 $\pm$ 0.116 &  1.398 $\pm$ 0.184 & 0.259 $\pm$ 0.114 & 47.28 $\pm$  3.13 & 3624.34 $\pm$  7.04 & 836.86 &    2.133 $\pm$    0.866 &  20 \\
77.7672.38 & 253.020 $\pm$ 0.166 &  2.103 $\pm$ 0.175 & 0.363 $\pm$ 0.166 & 258.54 $\pm$  9.82 & 3200.71 $\pm$ 11.93 & 660.92 &    5.154 $\pm$    1.676 &  19 \\
77.7673.25 & 252.165 $\pm$ 0.144 &  2.764 $\pm$ 0.215 & 0.201 $\pm$ 0.143 & 217.94 $\pm$  3.67 & 3248.27 $\pm$  4.43 & 392.27 &    8.067 $\pm$    2.019 &  19 \\
77.7910.59 & 319.326 $\pm$ 0.157 &  1.729 $\pm$ 0.168 & 0.300 $\pm$ 0.156 & 282.05 $\pm$  5.50 & 3181.29 $\pm$  7.37 & 473.72 &    2.203 $\pm$    0.725 &  18 \\
77.7912.33 & 322.424 $\pm$ 0.164 &  1.480 $\pm$ 0.183 & 0.257 $\pm$ 0.162 & 247.00 $\pm$  7.50 & 3111.40 $\pm$  5.92 & 320.79 &    0.973 $\pm$    0.383 &  19 \\
77.7912.36 & 300.304 $\pm$ 0.115 &  1.548 $\pm$ 0.147 & 0.618 $\pm$ 0.112 & 255.40 $\pm$  4.39 & 3145.92 $\pm$  4.22 & 519.20 &    0.970 $\pm$    0.428 &  19 \\
77.7912.66 & 264.590 $\pm$ 0.111 &  1.458 $\pm$ 0.178 & 0.487 $\pm$ 0.110 & 55.66 $\pm$  5.03 & 3136.40 $\pm$  3.33 & 339.35 &    0.725 $\pm$    0.307 &  19 \\
77.7914.39 & 237.481 $\pm$ 0.112 &  1.760 $\pm$ 0.130 & 0.120 $\pm$ 0.112 & 246.58 $\pm$  5.57 & 3421.15 $\pm$ 10.36 & 717.39 &    3.964 $\pm$    0.895 &  19 \\
77.8034.380 & 238.063 $\pm$ 0.253 &  2.140 $\pm$ 0.099 & 0.322 $\pm$ 0.229 & 207.27 $\pm$  5.82 & 3265.28 $\pm$ 10.60 & 401.59 &    3.459 $\pm$    0.981 &  19 \\
77.8035.48 & 302.518 $\pm$ 0.142 &  1.853 $\pm$ 0.167 & 0.081 $\pm$ 0.141 & 110.90 $\pm$  8.01 & 3186.59 $\pm$ 16.24 & 719.69 &    4.695 $\pm$    1.281 &  19 \\
\hline
\end{tabular}
\end{table*}

\section{Results}

In this section, we discuss some general results that are not model specific. Model specific results are presented in section~\ref{models}.

\subsection{The True Distribution of the Velocity Amplitude}
\label{vampdsn}

The distribution of velocity amplitude is an important parameter for most models of the sequence D phenomenon.

\begin{figure}
\begin{center}
\includegraphics[width=0.5\textwidth]{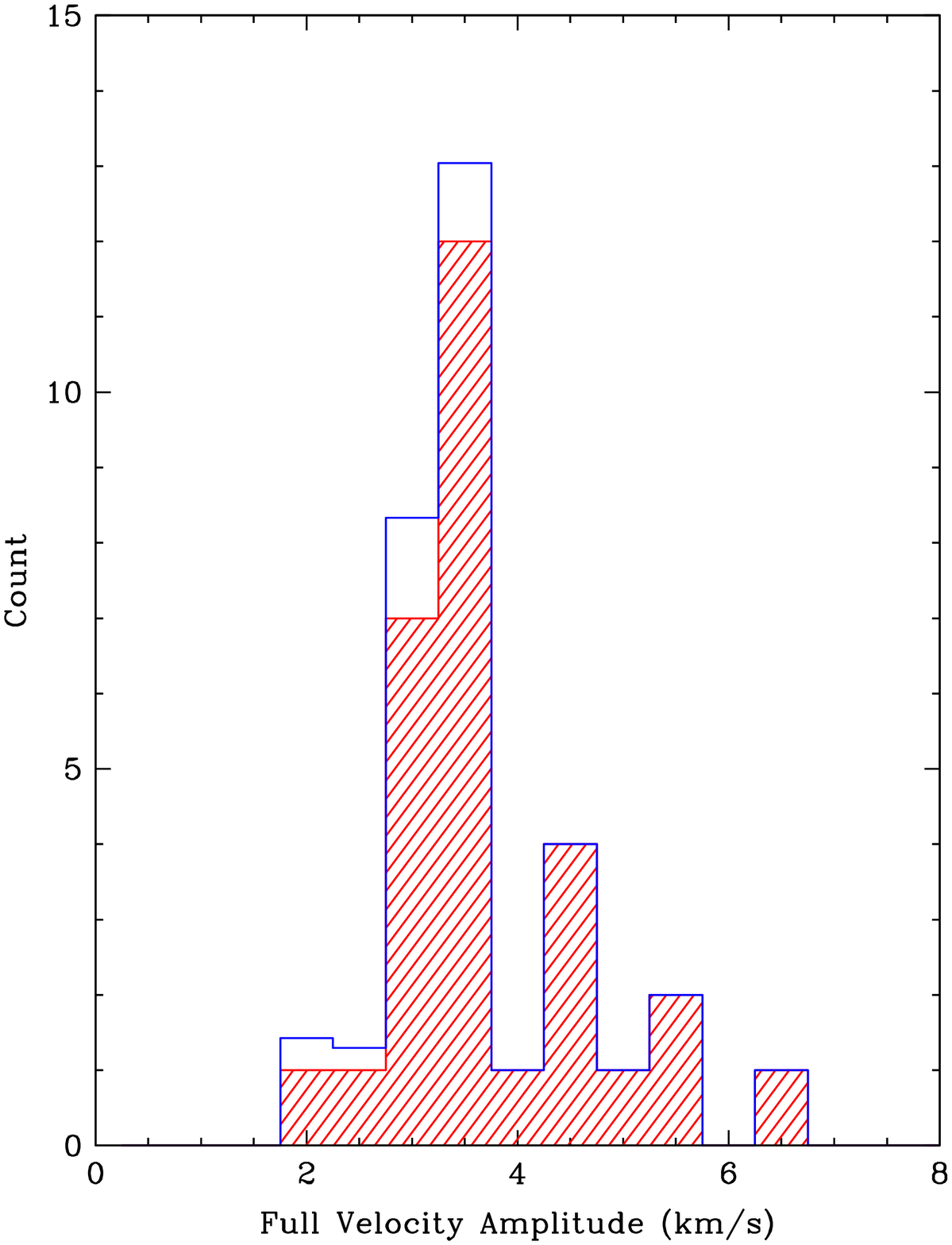}
\end{center}
\caption[Velocity Amplitude Histogram]{A histogram of the observed full velocity
  amplitude is plotted in red (forward shading). Note the clustering around $3.5 \rm{km\,s^{-1}}$. Overplotted in blue is the modified histogram to account for the observational detection probability calculated in our Monte Carlo simulation (see text for details).}
\label{vfamphist}
\end{figure}

A histogram of the velocity amplitudes of our sample is shown in Fig.~\ref{vfamphist}, with the values separated into bins of width
$0.5 \rm{km\,s^{-1}}$. It is important to note that we use the full or peak-to-peak amplitude, as opposed to the semi-amplitude. These amplitudes are obtained from the binary orbit fits to the observed velocities. We find that the velocity amplitude is concentrated mainly between 2.5
and $5.0 \rm{km\,s^{-1}}$, with the highest number of stars falling in the bin centred at $3.5 \rm{km\,s^{-1}}$. The distribution has a
median value of $3.53 \rm{km\,s^{-1}}$. Thus our sequence D stars all have low, relatively similar velocity amplitudes. These
results are consistent with the results of \cite{hinkle02}'s and \cite{sequenceDstars}'s analysis of Galactic LSP variables. 

There are very few stars in our sample with velocity amplitudes $<3.0 \rm{km\,s^{-1}}$. To find out if we could detect such stars we performed a Monte Carlo simulation. Fifty stars were simulated for each of the velocity amplitudes of 3.0, 2.0, 1.0 and $0.5 \rm{km\,s^{-1}}$. The period, eccentricity, and time of periastron were all selected randomly. The number of velocity points in each simulation was the same as the number of observations (19) and they were distributed in time identically to the observations. Random noise was added to each point, drawn from a normal distribution whose mean was the mean velocity error in the observations. In plots for visual examination, an errorbar was added for each point, drawn from a uniform distribution between 0.2 and $0.5 \rm{km\,s^{-1}}$(consistent with observed errors).

\begin{figure}
\begin{center}
\includegraphics[width=0.5\textwidth]{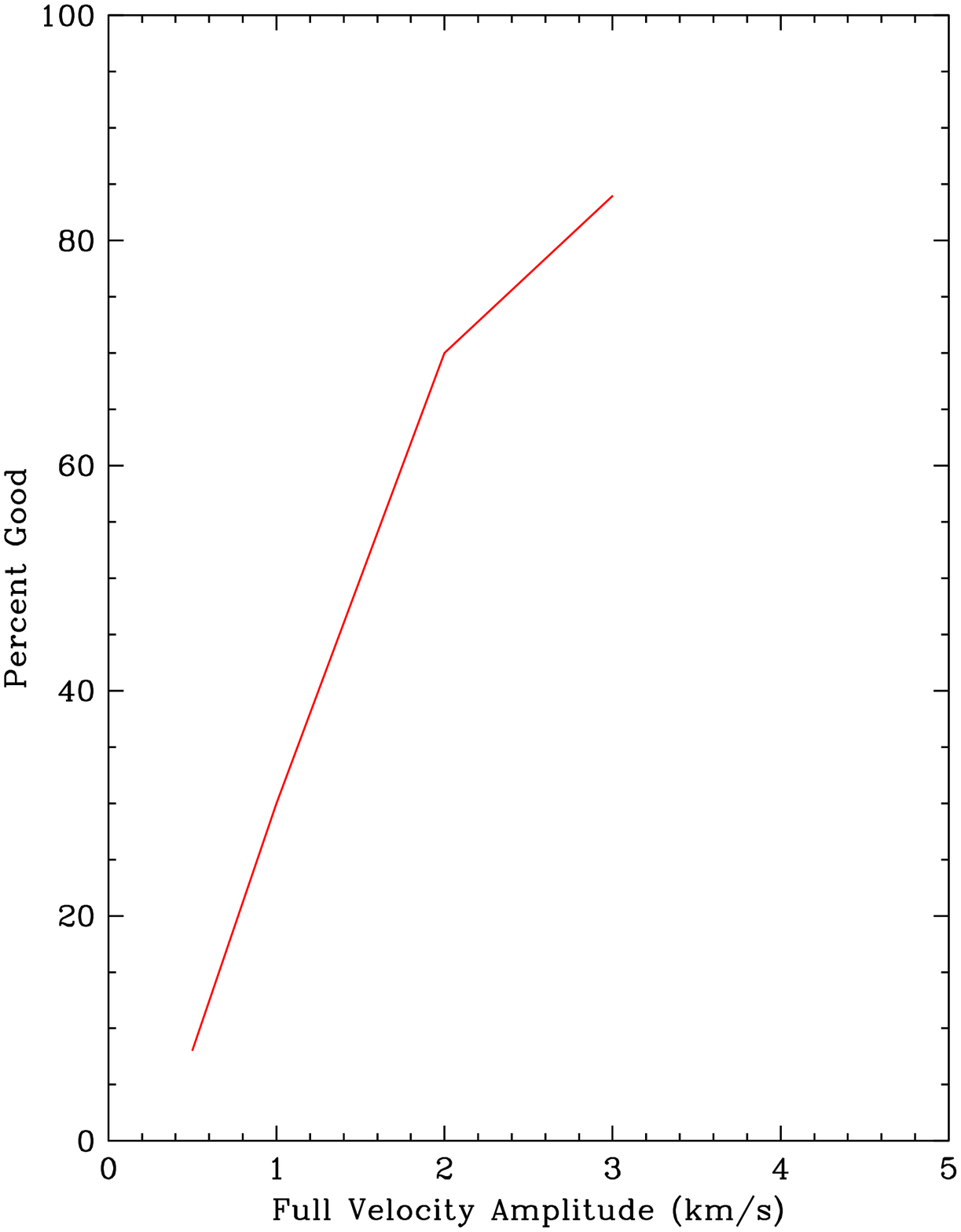}
\end{center}
\caption[Fraction of simulated velocity curves which pass our data reduction process, for different velocity amplitudes.]{Fraction of simulated velocity curves which pass our data reduction process, for different velocity amplitudes.}
\label{percentages}
\end{figure}

A binary fit was made to the simulated velocities in the same way as to the observed velocities (see section~\ref{obtaining}). The simulated data were put through the same reduction steps as the observed data, to see what percentage would pass. The percentages can be seen in Fig.~\ref{percentages}.
This tells us that we should be able to detect velocity amplitudes $<3.0 \rm{km\,s^{-1}}$. Using our simulated percentages, we can estimate the true number of stars at each velocity amplitude by dividing the observed histogram by the observable percentage. The resulting modified velocity amplitude histogram is shown overplotted on the observed histogram in Fig.~\ref{vfamphist}.

The fact that we do not observe many sequence D stars with velocity amplitudes $<3.0 \rm{km\,s^{-1}}$ suggests there are not many to be observed. We conclude that the true distribution of sequence D velocity amplitudes, represented in Fig.~\ref{vfamphist}, shows a strong peak near $3.5 \rm{km\,s^{-1}}$.

\subsection{Correlation of Light and Velocity Amplitudes} 

\begin{figure}
\begin{center}
\includegraphics[width=0.5\textwidth]{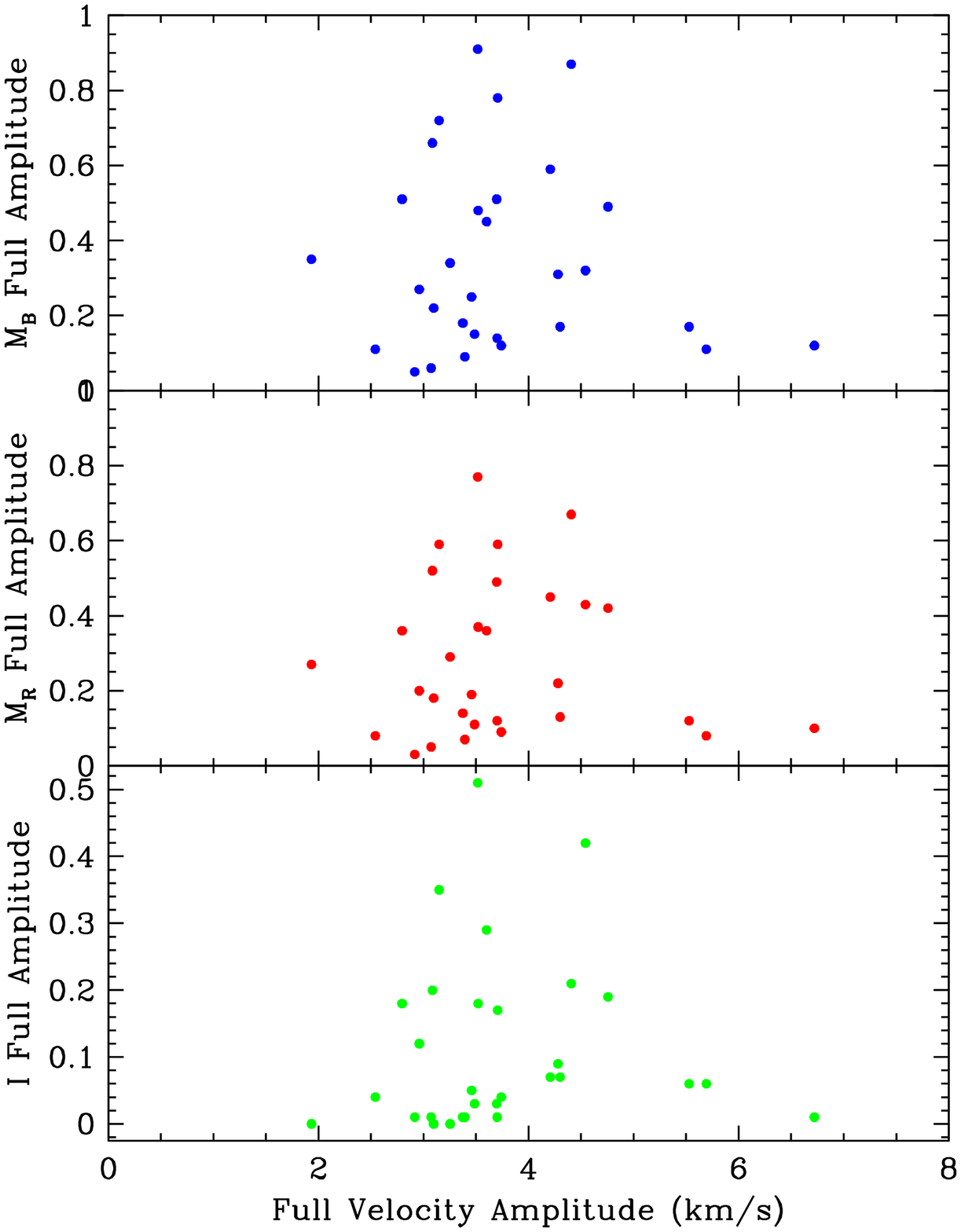}
\end{center}
\caption[Light amplitude plotted against velocity amplitude.]{Light
  amplitude plotted against velocity amplitude for $M_{B}$, $M_{R}$, and $I$.}
\label{v-light}
\end{figure}

Light amplitude is plotted against velocity amplitude in Fig.~\ref{v-light}. There is no obvious correlation between light and velocity amplitudes in
either $M_B$, $M_R$, or $I$. As noted previously the velocity amplitude has a small spread, and peaks at the small value of
$3.5 \rm{km\,s^{-1}}$. In particular, stars with small light amplitudes have velocity amplitudes as large as those stars with the largest light amplitudes.

\subsection{Correlation of Velocity Amplitude with Magnitude and Period}
\label{3.3}

\begin{figure}
\begin{center}
\includegraphics[width=0.5\textwidth]{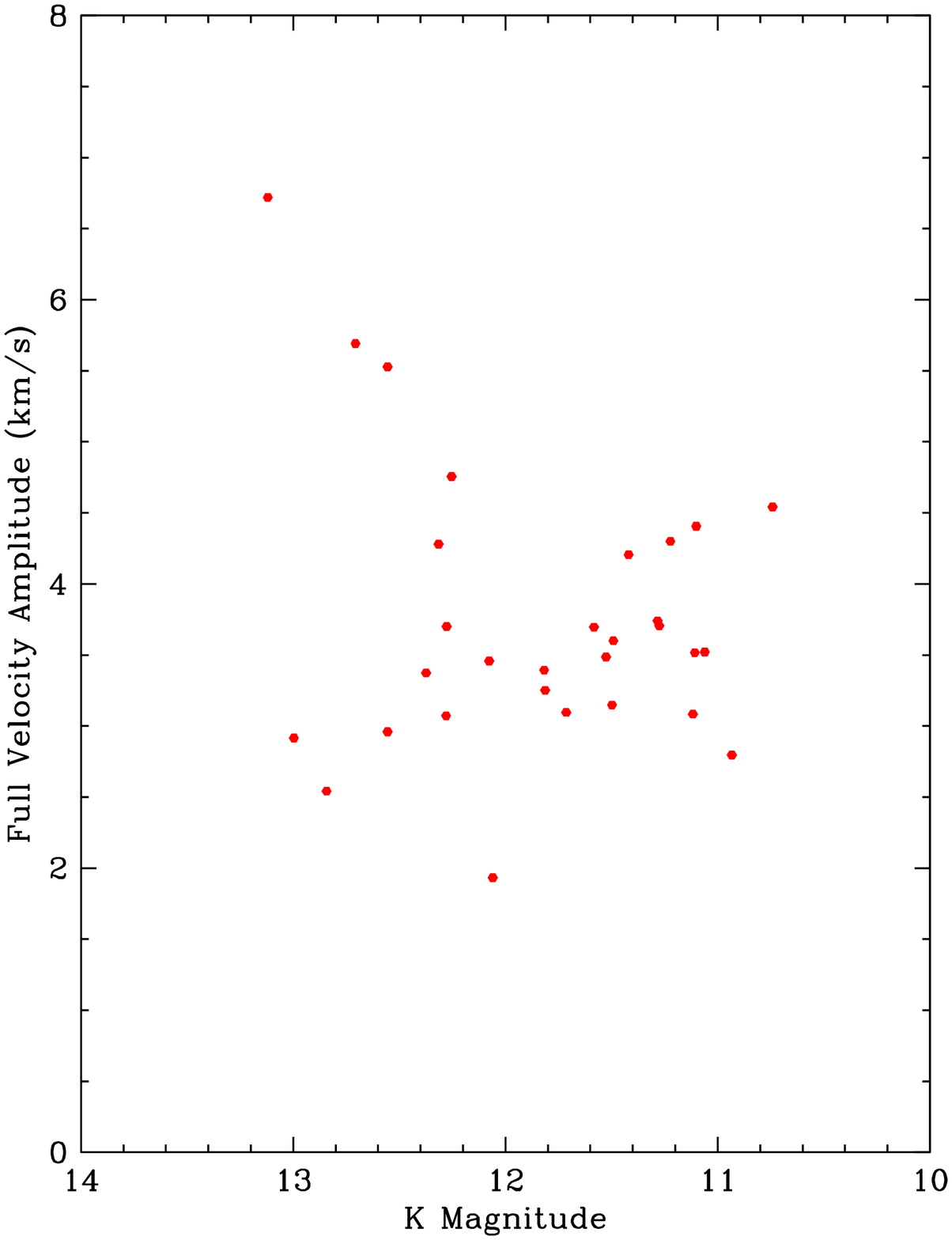}
\end{center}
\caption[K magnitude plotted against velocity amplitude.]{Velocity amplitude
  plotted against $K$ magnitude.} 
\label{k-v}
\end{figure}

\begin{figure}
\begin{center}
\includegraphics[width=0.5\textwidth]{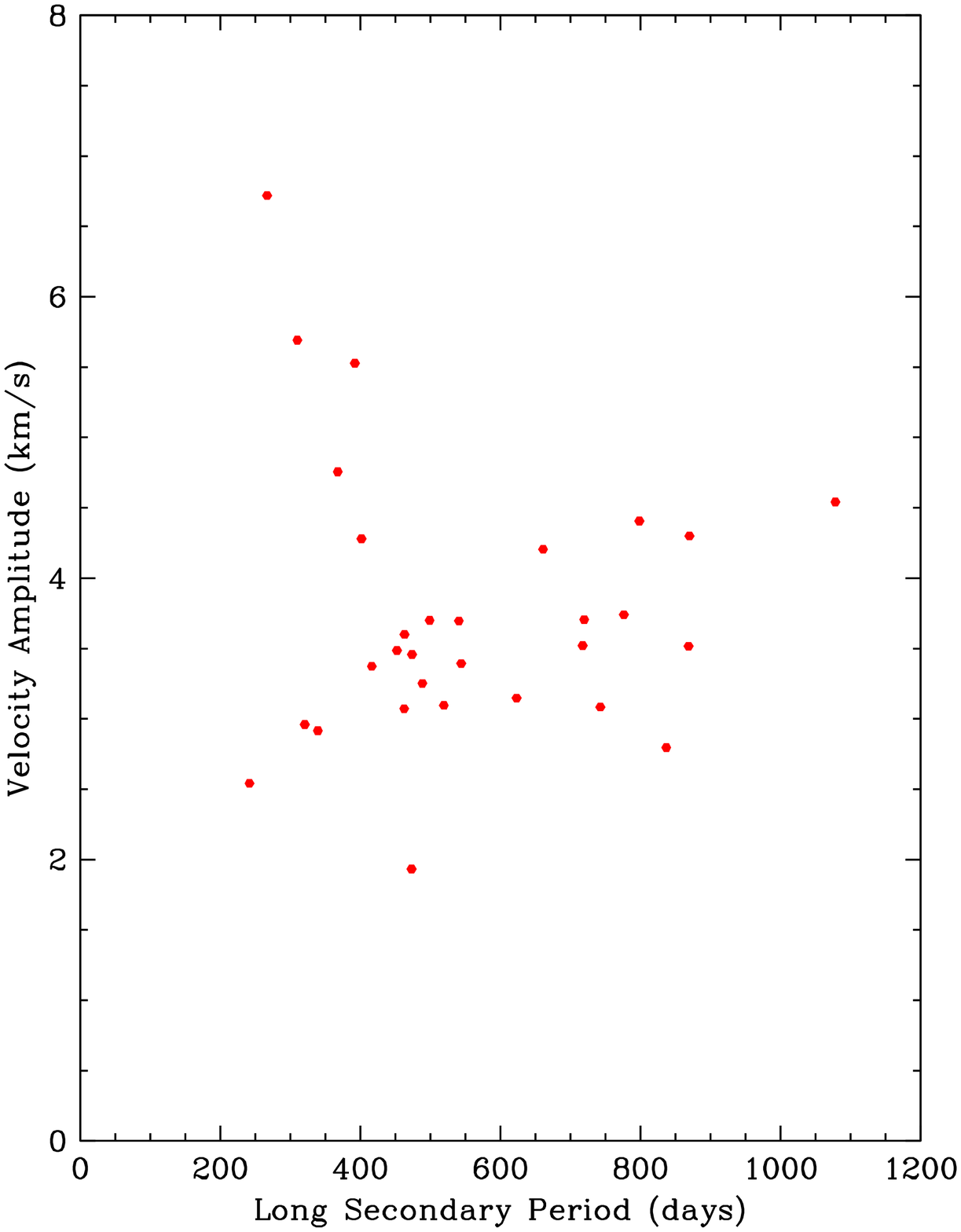}
\end{center}
\caption[LSP plotted against velocity amplitude.]{Velocity amplitude plotted against LSP.} 
\label{lsp-v}
\end{figure}

Velocity amplitude is plotted against $K$ magnitude in Fig.~\ref{k-v}, which shows a suggestion that velocity amplitude increases slowly with luminosity. A similar correlation is seen in Fig.~\ref{lsp-v} between velocity amplitude and LSP. However these possible correlations are disrupted by the presence of a few stars with larger velocity amplitudes, that fall apart from the general relation. 

\section{Testing and Constraining Models}
\label{models}

Here we examine some of the more plausible and popular models of LSPs, in the light of our new data. 

\subsection{The Binary Model}

The model of binary motion as an explanation for LSPs has been much debated since its proposal by \nocite{wood99mn} Wood et al.\ (1999). In
its favour, binary motion may be able to explain the sequence D velocity curves, and the light variation was originally thought by Wood et
al.\ to resemble an eclipse of the star by a dusty cloud surrounding a small, orbiting companion. However when this model was further examined by
\cite{hinkle02} and \cite{sequenceDstars}, two main problems were found: the companion mass was calculated to be always $\sim$0.1 $\rm{M_{\odot}}$, and
the characteristic shape of sequence D stars' velocity curves was shown to imply that the angle of periastron is not uniformly distributed
between $0$ and $2\pi$. With our larger sample we can place tighter constraints on these properties.  

\subsubsection{The Companion Mass}

In a binary system, the velocity amplitude and period give an estimate of the mass of the companion. For a typical sequence D star with an LSP of 500
days, a typical velocity amplitude of $3.5\ \rm{km\,s^{-1}}$, and an assumed total system mass of $M=1.5\ \rm{M_{\odot}}$, the orbital separation is of the order of 1.4 AU and the companion has a mass of $0.09\ \rm{M_{\odot}}$. This means it is a very low mass main sequence star or a brown dwarf. 
There is an observed deficit of brown dwarf companions to main sequence stars, compared to both more massive stellar companions and less massive planetary companions. This is known as the `Brown Dwarf Desert' \citep{mccarthy04, browndwarfdesert}. The range of likely companion masses can be found by calculating the companion mass for velocity amplitudes within one sigma of the mean of $3.5 \rm{km\,s^{-1}}$. We find a companion mass range of 0.06 to 0.12 $\rm{M_{\odot}}$. Using the fits to the data in Fig.~8 of \cite{browndwarfdesert}, we find that only 0.86 per cent of low mass main sequence stars have companions with masses in this range. Thus binarity is an unlikely model, when we remember that $\sim$30 per cent of low mass stars will exhibit LSPs when they pass through the AGB stage.

\subsubsection{Correlation of Velocity Amplitude with Light Amplitude, Magnitude and Period}

We would not expect the velocity and light amplitudes to be related in a binary model where the light variation is due to an eclipse phenomenon (by a cloud in the \nocite{wood99mn} Wood et al.\ 1999 model). However, these amplitudes may be related if the light variations were caused by distortion of the red giant by its unseen companion (see section~\ref{ellips}).  

As noted in section~\ref{3.3}, only a hint of a correlation is seen between velocity amplitude and $K$ magnitude (Fig.~\ref{k-v}), and with LSP (Fig.~\ref{lsp-v}). We cannot draw any conclusions from this in the context of a binary model. 

\subsubsection{The Distribution of the Angle of Periastron}

One of the most conclusive ways to test the plausibility of the binary model as a cause for the LSPs is to examine the distribution of the angle of
periastron, $\omega$. As \cite{hinkle02} and \cite{sequenceDstars} note, the observed distribution of $\omega$ should be uniform over the
whole range of angles, since one would expect binary orbits to be randomly aligned in space. 

\begin{figure}
\begin{center}
\includegraphics[width=0.5\textwidth]{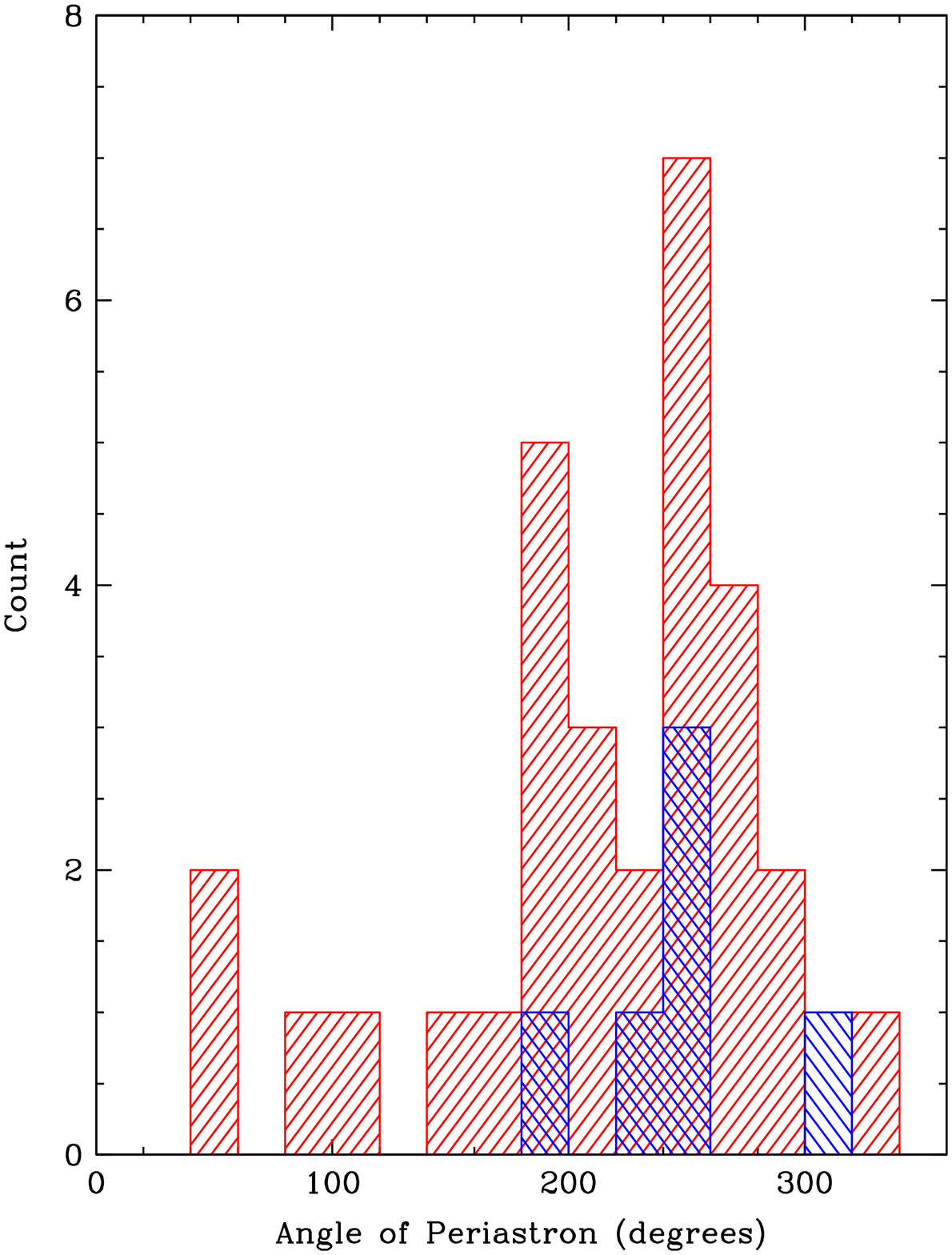}
\end{center}
\caption[Angle of Periastron Histogram]{The distribution of the angle of
  periastron, $\omega$. This study's sample is plotted in red (forward
  shading). The 
  combined sample of \cite{hinkle02} and \cite{sequenceDstars} is shown
  in blue (back shading). Two stars have been dropped from
  \citeauthor{hinkle02}'s sample and one from
  \citeauthor{sequenceDstars}'s, due to overly large errors for $\omega$.}
\label{omegahist}
\end{figure}

The angle of periastron was calculated by the program \textsc{Fitall} as part of the binary fit made to our radial velocity data. The
distribution of the angle of periastron is plotted as a histogram in Fig.~\ref{omegahist}, with the data divided into bins $20^{\circ}$ in width. Note that since the errors in the angle of periastron are typically around $5^{\circ}$ (Table~\ref{orbital}), the distribution of angle of periastron shown in Fig.~\ref{omegahist} will not be significantly broadened by these errors. There is a clear bias in the distribution towards angles $>180^{\circ}$, and the median of the sample is $227^{\circ}$. The distribution of the angles of periastron calculated by \cite{hinkle02} and \cite{sequenceDstars} for their small samples of Galactic stars has been overplotted. It is clear that their values of $\omega$
lie in the same region as the bulk of the present sample. This shows that the LSPs share a common cause, whether found in the LMC or in our own Galaxy.

\cite{hinkle02} and \cite{sequenceDstars} both found that the characteristic shape of the velocity curves suggests an eccentric orbit and a large angle of periastron, and that the distribution of the calculated angle of periastron was inconsistent with what would be expected if the sequence D stars were
binaries, as no satisfactory explanation could be found for this bias towards higher angles. But given their small sample size, their
conclusions were not highly significant. We can quantify and greatly improve these claims by using a Kolmogorov--Smirnov, or K--S Test, with our new data. 

We have used the one-sample K--S Test, in the form of the Fortran 77 subroutine \textsc{ksone} \citep[provided in][]{numerical}, to compare our
distribution of the angle of periastron to the uniform distribution. We find that the probability that our distribution of $\omega$ is
consistent with the uniform distribution is $1.4\times10^{-3}$. In other words, the probability that the sequence D stars are binaries is
extremely small. This is a major result from this study. 

An angle of periastron between $180^{\circ}$ and $360^{\circ}$ means that at periastron, the red giant is closest to the observer, with the smaller companion further away. It is hard to see what selection effects could cause an LSP to be observed only in these binary orientations.

\subsubsection{The Distribution of the Eccentricity}

\begin{figure}
\begin{center}
\includegraphics[width=0.5\textwidth]{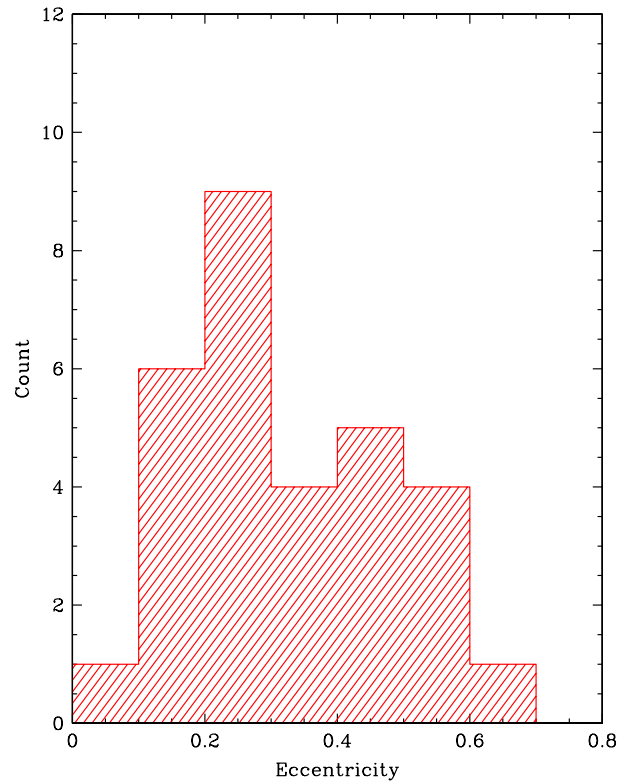}
\end{center}
\caption[Eccentricity Histogram]{The distribution of the eccentricity.}
\label{ehist}
\end{figure}

A histogram showing the distribution of the eccentricity is shown in Fig.~\ref{ehist}. The eccentricity was calculated by \textsc{Fitall} as one of the orbital parameters of the binary fit to the velocity data.

The plot shows that if the sequence D stars are caused by binary motion, then they are in eccentric orbits. This confirms the suggestion of \cite{hinkle02} and \cite{sequenceDstars}. The median eccentricity of the sample is 0.3.

\subsubsection{Ellipsoidal Variables}
\label{ellips}

An ellipsoidal variable is a binary star which is tidally distorted by an orbiting companion into an ellipsoidal shape. Light variability is
caused mostly by the variation of the apparent surface area of the star seen by the observer as the star rotates. Its velocity curve is dominated by
orbital motion, with a small contribution from the rotation of the limb-darkened ellipsoid. 

The light curve for an ellipsoidal variable is expected to have two maxima and minima per orbit, due to the star's shape. However, the velocity curve is expected to have only one maximum and minimum per orbit. We can use this characteristic to investigate the plausibility of ellipsoidal variability as an explanation for the LSP variation.

\nocite{ogleellipsoidalmn} Soszy\'nski et al. (2004b) show that the light curves of \emph{sequence E} variables are satisfactorily explained by ellipsoidal variability. Their
data show a partial overlap of sequences D and E in the P--L diagram, and so they suggest that the Long Secondary Periods may have a binary
ellipsoidal origin. \cite{soszynski07} explores this possibility further by searching for ellipsoidal or eclipsing shapes in residual LSP light
curves. He finds these variations in $\sim$5 per cent of his sample, and adopts the binary model to explain LSPs. 

However, with our new dataset (which includes some sequence E stars), the difference between sequence D and sequence E mechanisms can now be
vividly demonstrated by the difference between the light and velocity curves of sequence D and E stars. The phased-up light and velocity
curves of sequence E stars show that the light curve completes two full cycles to every single cycle of the radial velocity, as expected for an
ellipsoidal variable \nocite{sequenceEstars} \citep[][ Nicholls, Wood \& Cioni, in preparation]{betsy}. However, Fig.~\ref{fitting} shows that sequence D stars do not display this behaviour: the phased-up light and velocity curves match each other cycle for cycle. Furthermore, sequence E stars typically have velocity amplitudes of $30 \rm{km\,s^{-1}}$ or more while the sequence D stars have significantly lower typical amplitudes of $3.5 \rm{km\,s^{-1}}$. 
\cite{soszynski07} predicted that sequence D stars showing residual ellipsoidal or eclipsing-type variations should have large velocity amplitudes,
similar to those observed for sequence E stars. One of our stars (77.7671.282) was noted by \citeauthor{soszynski07} as a double-humped LSPV and its
velocity amplitude is only $\sim$4 $\rm{km\,s^{-1}}$. Additionally, none of our LSP sample show typical sequence E velocity amplitudes (see Fig.~\ref{vfamphist}). Sequence D and E variables clearly have distinctly different velocity amplitudes.

There is an unambiguous distinction between the mechanisms of variability responsible for causing sequences D and E\@. We can now
state with some certainty that binary star ellipsoidal variability is not the cause of the LSP\@. 

Similar to the sequence E ellipsoidal variables are symbiotic binaries, which are red giants with an accreting white dwarf companion, often in an eccentric orbit. Radial velocity amplitudes of these systems are usually $\geq8 \rm{km\,s^{-1}}$, and the red giant may be a Mira variable \citep{symbioticmira}. However, symbiotic spectra show high temperature emission lines from accretion onto the white dwarf, something that has not been seen in either the sequence E or sequence D spectra. There does not seem to be any relation between the symbiotic stars and the sequence D stars.

\begin{figure}
\begin{center}
\includegraphics[width=0.5\textwidth]{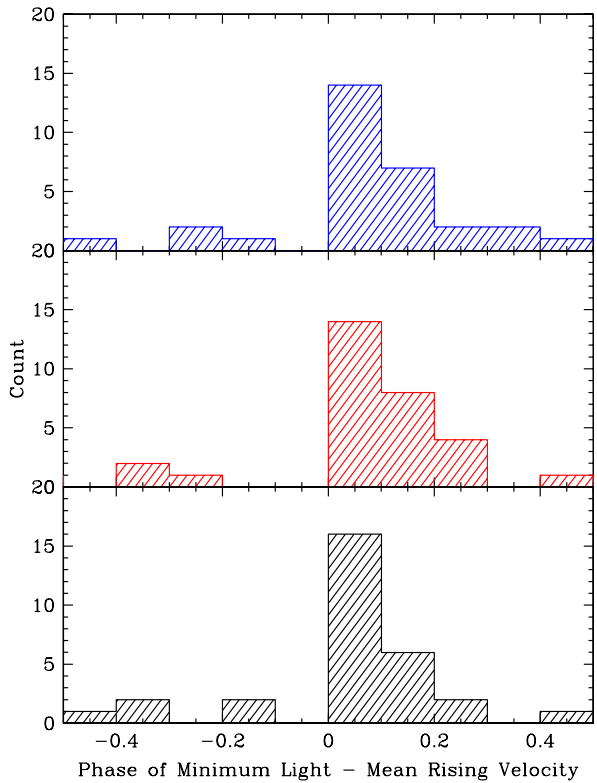}
\end{center}
\caption[Histogram of Phase Difference between LSP Light and
  Velocity]{Histograms of the difference between the phase of LSP minimum
  light and the phase of mean rising LSP radial velocity. The top panel shows
  the phase difference for MACHO $M_{B}$, the middle panel $M_{R}$, and
  the bottom panel OGLE $I$. Radial velocity is defined with respect to
  the observer: a positive radial velocity corresponds to matter moving away from the observer i.e.
  infalling to the centre of the star in a radial pulsation model.}
\label{phasehist}
\end{figure}

\subsection{Radial Pulsation}
\label{radial}

There are several ways in which the present dataset can be used to test radial pulsation models:

\subsubsection{The Phase Difference between Light and Velocity Curves}
\label{phase}

Minimum light and the mean of the velocity during increase are relatively well defined for our sequence D stars. In order to examine their relative phases, a histogram of the phase difference between minimum light and mean rising velocity is shown in Fig.~\ref{phasehist}. The phases of minimum light and mean rising velocity were obtained from the fits made to the light and velocity curves. Each panel shows a strong peak of phase differences lying between 0 and 0.2 and there is little spread in phase. 

\cite{lebzelter00} and \cite{lebzelter} find that for radially pulsating SRVs, the light lags the radial velocity by roughly 0.5 in phase, which means that minimum light occurs $\sim$0.25 in phase after the time of mean rising velocity. Our model SRV (Fig.~\ref{sr_model}) shows that minimum light occurs close to 0.23 in phase after mean rising velocity. Our sequence D stars are significantly different. Taking the main peak of light-velocity phase differences in Fig.~\ref{phasehist} (values between $-0.1$ and $0.3$), the mean phase shift is $0.1$, and the standard deviation is $0.08$. More simply, minimum light usually occurs $\sim$0.1 in phase after mean rising velocity. Radial pulsation therefore may not be consistent with this aspect of the LSP phenomenon, although this is not a strong conclusion.

\subsubsection{The Shape of the Velocity Curve}

During large amplitude radial stellar pulsation (as in Mira variables), a shock wave from the interior of the star causes the surface layers
to rapidly accelerate outwards, and then settle back more slowly. This is also shown in the shape of the velocity curve of our low amplitude model SRV, Fig.~\ref{sr_model}. The velocity curves of most of our sequence D stars show exactly the opposite of this behaviour: typically we find a rapid increase in radial velocity with a correspondingly slower decline (see Fig.~\ref{fitting}). In a radial pulsation model, this translates to a slow increase in stellar
radius and a quick decrease. This has not been observed in any red giant stars known to be radially pulsating variables \citep[e.g.][ for SR variables and Mira variables respectively]{lebzelter99, hinkle84}. 

\subsubsection{Correlation of Velocity Amplitude with Light Amplitude, Magnitude and Period}

The lack of correlation between velocity amplitude and light amplitude for the stars in our sample (see Fig.~\ref{v-light}) is an unexpected result for pulsation mechanisms as they would be expected to show a positive correlation between light and velocity amplitude \citep{hinkle97}. Additionally, correlations of velocity amplitude with luminosity and with period are expected for radially pulsating variables. However Figs.~\ref{k-v} and~\ref{lsp-v} show that any correlations between these quantities in our sample are vague at best. We are unable to draw any conclusions from this in the context of radial pulsation.

\subsubsection{The Variation of Stellar Radius}
\label{rvar}

The stellar radius can be calculated from our data in several ways. Our first method was to calculate $R$ using the Stefan-Boltzmann equation 
$L=4\pi R^2\sigma T^4_{\rm{eff}}$, with $L$ and $T_{\rm{eff}}$ calculated from MACHO photometry (see Appendix).

\begin{figure*}
  \begin{center}
  \includegraphics[width=0.9\textwidth]{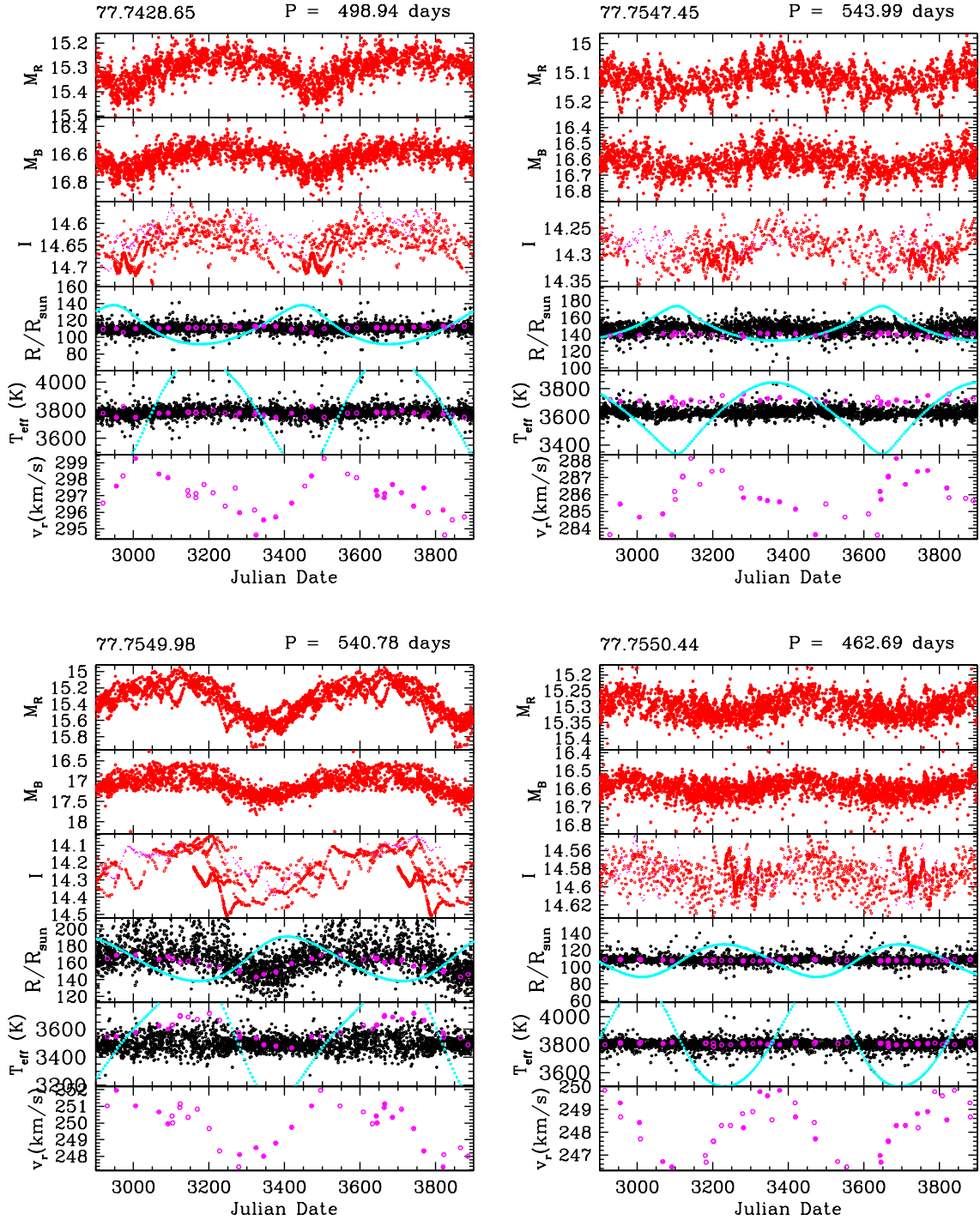}
  \end{center}
  \caption[Different properties of a typical sequence D star plotted
  against Julian Date.]{Different properties of four typical sequence D
  stars plotted against Julian Date. Solid symbols denote values plotted
  at the date of measurement while open symbols are values that have
  been shifted forward or backward by one or more periods. \emph{Top
    panel}: MACHO Red light curve. \emph{Second panel}: MACHO Blue light
  curve. \emph{Third panel}: OGLE $I$ light curve. \emph{Fourth panel}:
  Variation of $R_{\rm{phot}}$ (small black points), variation of
  $R_{\rm{vel}}$ (cyan curve) and variation of $R_{\rm{spec}}$ (large
  magenta points). \emph{Fifth panel}: Variation of $T_{\rm{eff(phot)}}$
  (small black points), variation of $T_{\rm{eff(spec)}}$ (large magenta
  points) and variation of $T_{\rm{eff(vel)}}$ (cyan
  curve). \emph{Bottom panel}: Observed radial velocity. See text for details. Plots for the entire sample are available in the online version of this paper.}  
  \label{big}
  \end{figure*}

Examples of the variation of stellar radius calculated from the photometry, $R_{\rm{phot}}$, can be seen in Fig.~\ref{big}. There is generally
a clear periodic curve at the period of the LSP as well as a scatter due to the primary pulsation. 

Variation of stellar radius can also be calculated from the velocity, using $\Delta R=\int v dt$ where $v$ is the fit to the velocity data made by
\textsc{Fitall}. If we assume that the velocity changes are due to radial pulsation, then this gives the expected radius variation for
radial pulsation, $R_{\rm{vel}}$, where the resulting radius has been normalised to the median photometric radius. Here the velocity fit data
was scaled by a factor of 1.4 to convert from observed to true pulsation velocity in red giants \citep{scholzwood, sequenceDstars}. The radius change
computed from the velocity fit is shown in Fig.~\ref{big}. Again it shows a clear periodic curve. 

It is also possible to calculate stellar radius from the Stefan-Boltzmann equation using an effective temperature derived from the
spectra, as opposed to the photometry (spectral $T_{\rm{eff}}$ derivation is described in section~\ref{teffvar}). The luminosities at the times of the
spectra were in this case calculated from the fits to $M_B$, $M_R$ and ($M_B-M_R$) as described in the Appendix. The radius calculated in this manner, $R_{\rm{spec}}$, can also be seen in Fig.~\ref{big}. 

\begin{figure}
\begin{center}
\includegraphics[width=0.5\textwidth]{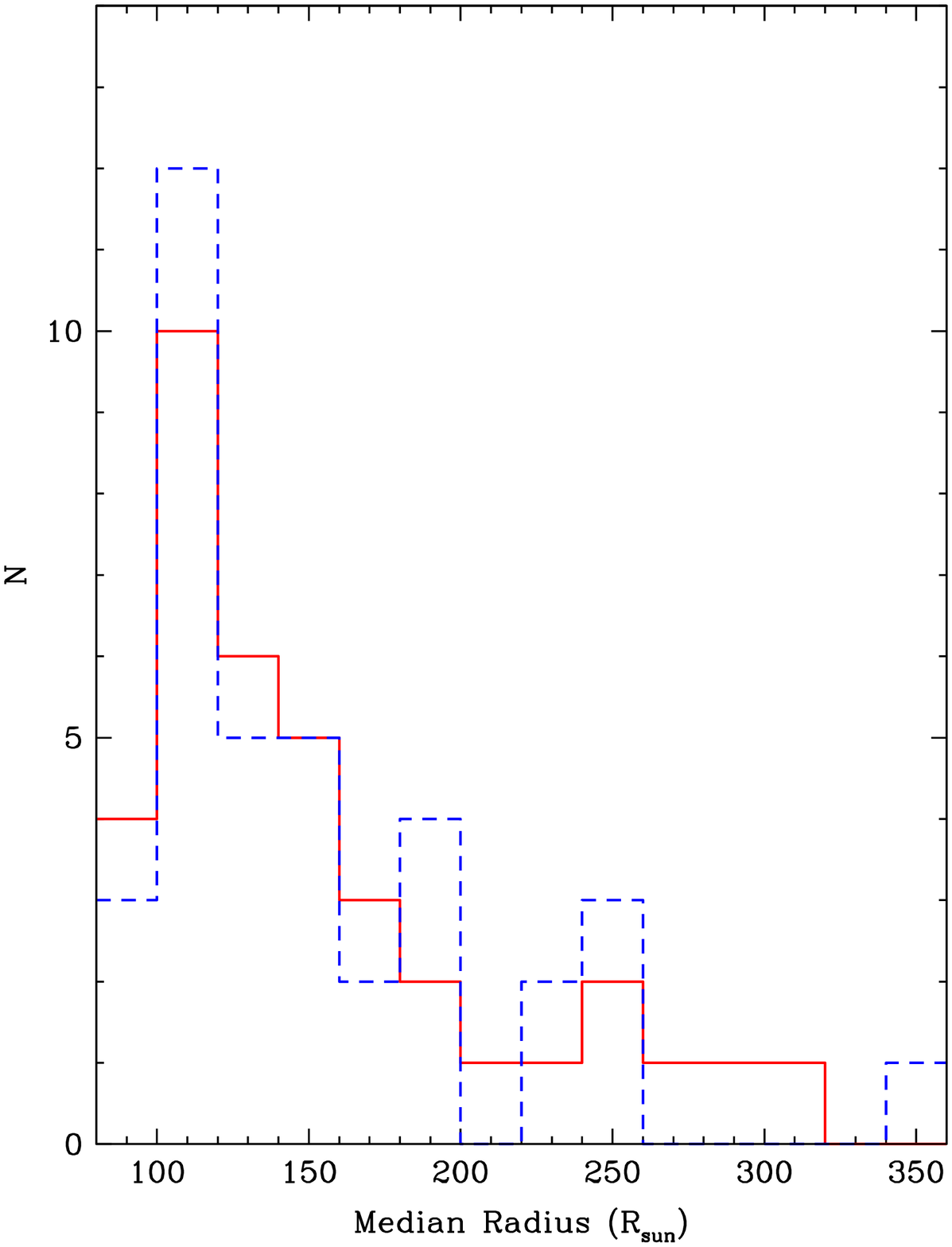}
\end{center}
\caption[Distribution of median values of different radius variations.]{Distribution of median values of $R_{\rm{phot}}$ (solid red line) and $R_{\rm{spec}}$ (dashed blue line).}  
\label{rhist}
\end{figure}

A histogram comparing the median values of $R_{\rm{phot}}$ and $R_{\rm{spec}}$ is shown in Fig.~\ref{rhist}. For both radius estimates, most stars have median radii lying between 100 and 200 $\rm{R_{\odot}}$. The median values for the entire sample for $R_{\rm{phot}}$ and $R_{\rm{spec}}$ are 135.4 and $132.1 \rm{R_{\odot}}$. $R_{\rm{vel}}$ is not shown, as it is normalised to $R_{\rm{phot}}$.

The radial amplitudes of our stars mostly lie between 3 and 60 $\rm{R_{\odot}}$ for all three variations. The median amplitudes of the sample for $\Delta R_{\rm{phot}}$, $\Delta R_{\rm{spec}}$ and $\Delta R_{\rm{vel}}$ are 5.92, 5.62, and $40.02 \rm{R_{\odot}}$ respectively. 

\begin{figure}
\begin{center}
\includegraphics[width=0.5\textwidth]{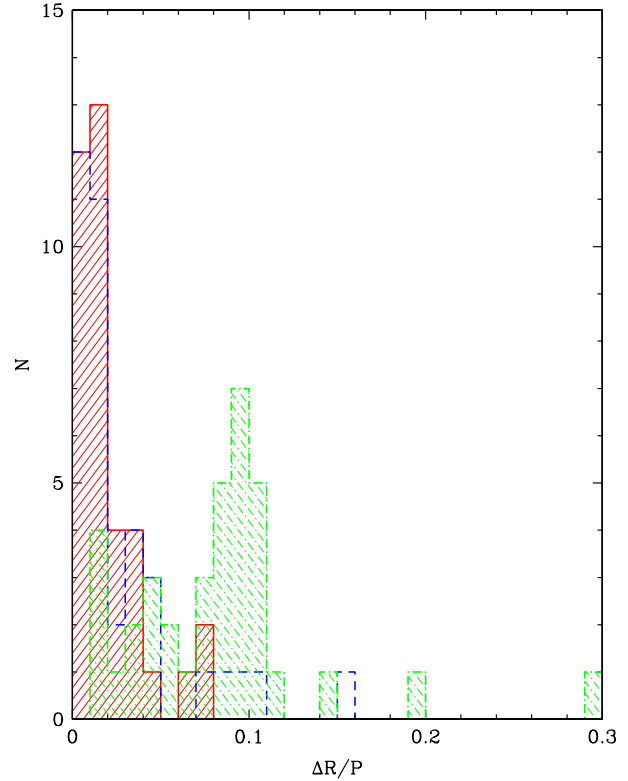}
\end{center}
\caption[Distribution of radial amplitude/period.]{Distribution of radial amplitude/period for $R_{\rm{phot}}$ (solid red line, forward shading), $R_{\rm{spec}}$ (dashed blue line) and $R_{\rm{vel}}$ (dot-dashed green line, back shading).}  
\label{ramplhist}
\end{figure}

Fig.~\ref{ramplhist} shows the distribution of radial amplitude $\Delta R$ divided by period, $P$. This quantity is a more sensitive indicator of differences between stars and models than the amplitude itself. In a radially pulsating model, one would expect the amplitude, $\Delta R$, to be tightly correlated to the period if the velocity amplitudes are the same i.e. $\Delta R$/$P$ would be a constant. Given that our stars show very similar velocity amplitudes (see Fig.~\ref{vfamphist}) we should expect tight, and similar, peaks for all three radius variations.

From Fig.~\ref{ramplhist} it is clear that the distributions of $\Delta R_{\rm{phot}}$/$P$ and $\Delta R_{\rm{spec}}$/$P$, are similar, but they are not the same as for $\Delta R_{\rm{vel}}$/$P$. $\Delta R_{\rm{vel}}$ is computed assuming radial pulsation. Given that it appears that the three radius variations do not agree, the assumption that $\Delta R$ is due to radial pulsation of the star is unlikely.

\begin{figure}
\begin{center}
\includegraphics[width=0.5\textwidth]{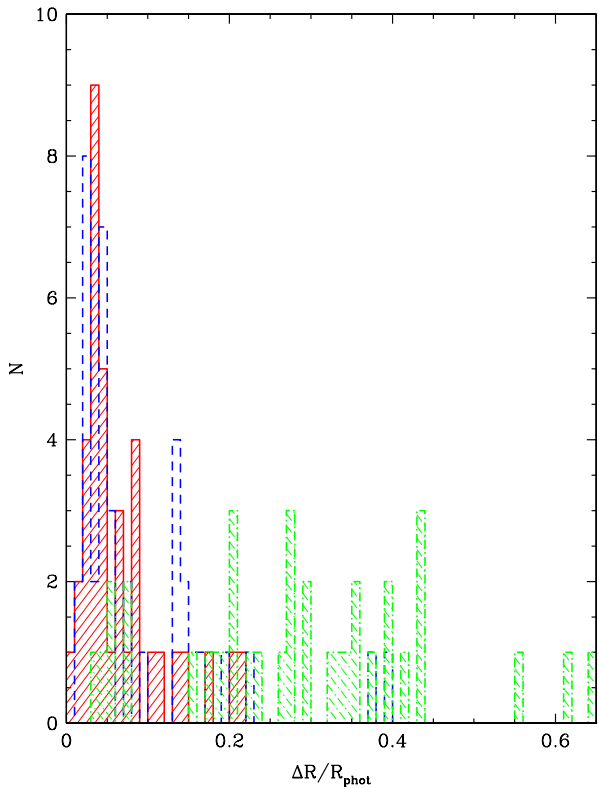}
\end{center}
\caption[Distribution of radial amplitude/radius.]{Distribution of relative radial amplitude for $R_{\rm{phot}}$ (solid red line, forward shading), $R_{\rm{spec}}$ (dashed blue line) and $R_{\rm{vel}}$ (dot-dashed green line, back shading).}  
\label{ramplrphot}
\end{figure}

Fig.~\ref{ramplrphot} shows the distribution of $\Delta R$ divided by $R$. For $R_{\rm{phot}}$ and $R_{\rm{spec}}$, the fractional radius change is usually 10 per cent or less. For $R_{\rm{vel}}$, the fractional radius change is much larger, usually between 10 and 50 per cent. Once again, the assumption that the measured radial velocity variations are due to radial pulsation does not seem to be correct.

\begin{figure}
\begin{center}
\includegraphics[width=0.5\textwidth]{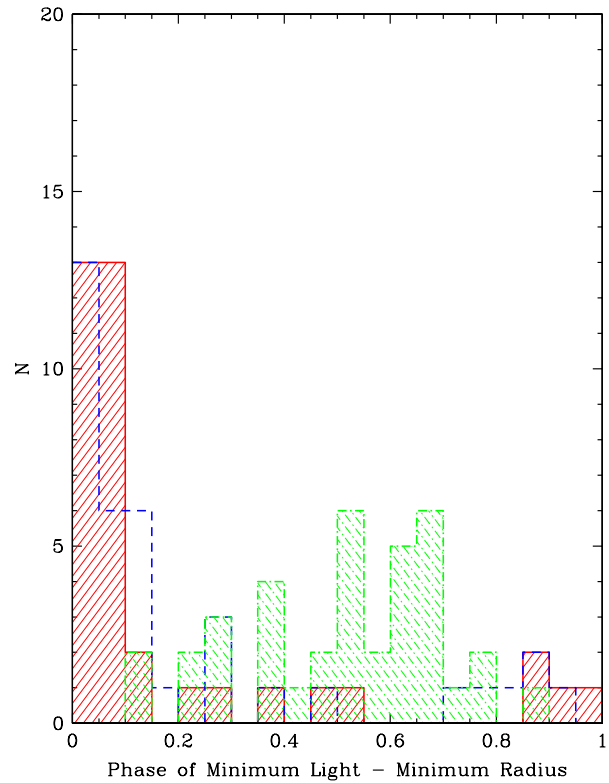}
\end{center}
\caption[Distribution of relative phases of minimum light and minimum radius.]{Distribution of phase of minimum light - phase of minimum radius for $R_{\rm{phot}}$ (solid red line, forward shading), $R_{\rm{spec}}$ (dashed blue line) and $R_{\rm{vel}}$ (dot-dashed green line, back shading).}  
\label{rphasehist}
\end{figure}

A histogram comparing the phases of the three radius variations to the phase of the $M_R$ light curve is shown in Fig.~\ref{rphasehist}. We compared the phase of minimum radius with the phase of minimum light, as these are both well-defined parts of the curves. In order to calculate these phases a Fourier series with four terms was fit to the radius curves. For the light curve, the previously-used Fourier fit was used (see Fig.~\ref{fitting}). Fig.~\ref{rphasehist} shows that $R_{\rm{phot}}$ and $R_{\rm{spec}}$ both vary almost in phase with the light variation (phase differences generally $\leq$0.1). On the other hand, $R_{\rm{vel}}$ is more likely to be out of phase with the light (and consequently with the other radius variation estimates). Again, the assumption that the observed radius variations can be attributed to radial pulsation leads to inconsistencies.

Using our model SRV (see section~\ref{correcting}) minimum radius occurs about 0.7 in phase before minimum light (Fig.~\ref{sr_model}). Of our variations, $R_{\rm{vel}}$ is the most similar to this. $R_{\rm{phot}}$ and $R_{\rm{spec}}$, whose phases agree well with each other, do not show the radius--light phase relation expected for radial pulsation.

Overall, the discrepancy between the variations of $R_{\rm{vel}}$ and the other two radius estimates suggests that radial pulsation is unlikely to be the cause of the LSPs.

\subsubsection{The Variation of Effective Temperature}
\label{teffvar}

The variation of $T_{\rm{eff}}$ was calculated from the photometry, from the spectra and from the velocity fit. 

The variation in the photometric $T_{\rm{eff}}$ ($T_{\rm{eff(phot)}}$), whose derivation is described in the appendix, is shown in Fig.~\ref{big}.

$T_{\rm{eff}}$ was also calculated using the depth of the TiO bandhead at 7054 \AA\ ($T_{\rm{eff(spec)}}$). The depth was calculated as the
ratio of the mean count level of a 4 \AA\ band longward of the bandhead to the mean count level of a 4 \AA\ band shortward of the bandhead,
taking into account the fact that the LMC has an average velocity of $270 \rm{km\,s^{-1}}$, leading to a longwards wavelength shift of
6.3 \AA\ in the spectra. 

In order to calculate the effective temperature from the TiO depth we calculated the depth of the TiO bandhead for 9 Wing Spectral Standard
stars. Then a quadratic fit was made to the TiO Depth--Spectral Type data. This quadratic fit was used to calculate the spectral types
of our sequence D stars from their TiO depths, and the effective temperatures were then calculated using the \cite{ridgway} $T_{\rm{eff}}$--Spectral type relation. An example of the typical variation of $T_{\rm{eff(spec)}}$ in these stars can be seen in Fig.~\ref{big}. 

Some of the warmer sequence D stars in our sample showed little to no evidence of TiO in their spectra, and so we were unable to include
these in the $T_{\rm{eff(spec)}}$ calculation. However, good-quality velocity curves were not necessary for calculating $T_{\rm{eff(spec)}}$ so a number of stars previously discounted from the velocity studies were added in where consideration of radius and effective temperature was required. Overall there was a net increase in sample size from 30 to 37 (the stars added in were excluded only from the velocity amplitude, angle of periastron, eccentricity and velocity phase parts of this work).

We also calculated the effective temperature variation expected for radial pulsation, $T_{\rm{eff(vel)}}$. This was calculated from the radius
change computed for pulsation in section~\ref{rvar}, the luminosity calculated from the fits to the photometry, and the Stefan-Boltzmann equation. The variation of $T_{\rm{eff(vel)}}$ can be seen in Fig.~\ref{big}.

\begin{figure}
\begin{center}
\includegraphics[width=0.5\textwidth]{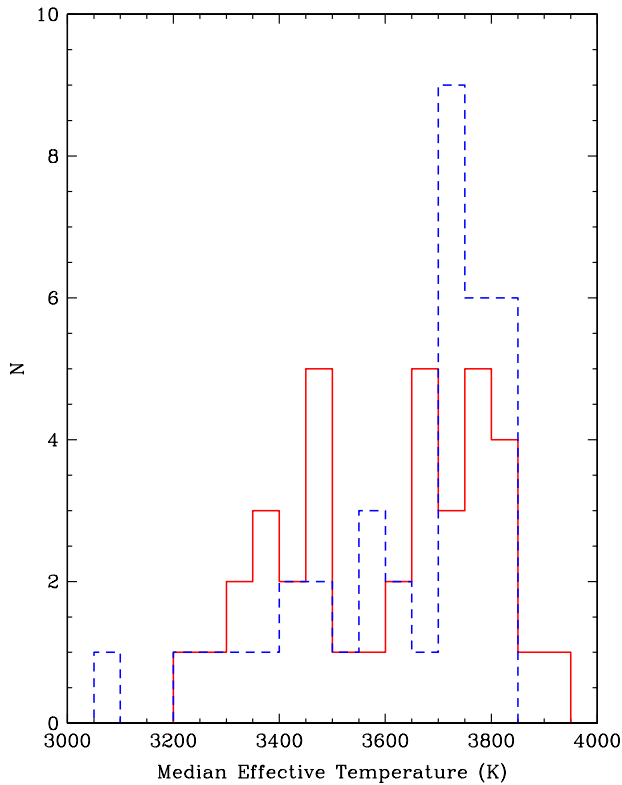}
\end{center}
\caption[Distribution of effective temperature.]{Distribution of median values of $T_{\rm{eff(phot)}}$ (solid red line) and $T_{\rm{eff(spec)}}$ (dashed blue line).}  
\label{thist}
\end{figure}

The distributions of the median values for each star of $T_{\rm{eff(phot)}}$ and $T_{\rm{eff(spec)}}$ are shown in Fig.~\ref{thist}. Most stars have median effective temperatures lying between 3300 and $3800 K$. The median values for the entire sample for $T_{\rm{eff(phot)}}$ and $T_{\rm{eff(spec)}}$ are 3662.9 and $3725.2 K$. $T_{\rm{eff(vel)}}$ is not shown, as it is calculated from $R_{\rm{vel}}$, which is normalised to $R_{\rm{phot}}$.

\begin{figure}
\begin{center}
\includegraphics[width=0.5\textwidth]{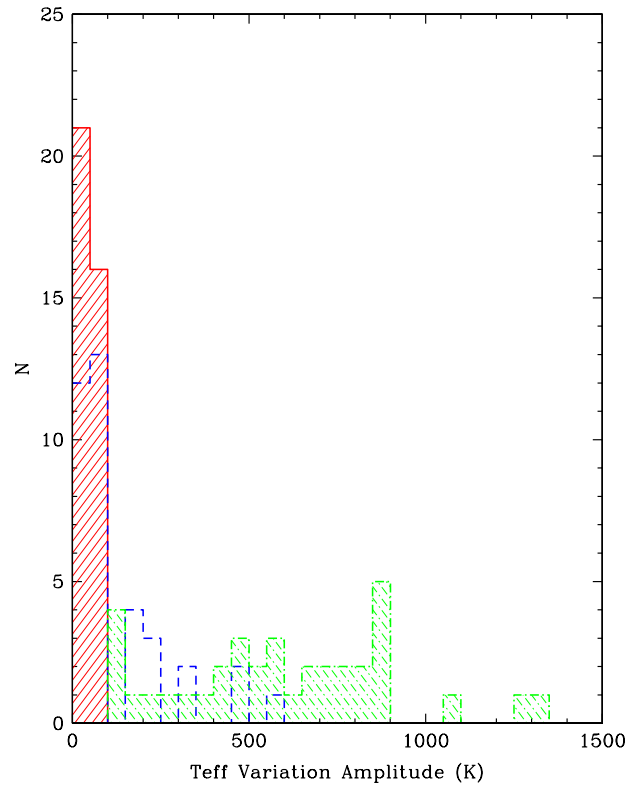}
\end{center}
\caption[Distribution of effective temperature amplitude.]{Distribution of amplitude of effective temperature variation for $T_{\rm{eff(phot)}}$ (solid red line, forward shading), $T_{\rm{eff(spec)}}$ (dashed blue line) and $T_{\rm{eff(vel)}}$ (dot-dashed green line, back shading). The higher end of the $\Delta T_{\rm{eff(vel)}}$ distribution has been cut off, in order to show greater detail at the peaks of $\Delta T_{\rm{eff(phot)}}$ and $\Delta T_{\rm{eff(spec)}}$.}  
\label{tamplhist}
\end{figure}

Fig.~\ref{tamplhist} shows the distribution of $T_{\rm{eff}}$ amplitude for the three variations. The amplitude distributions of $T_{\rm{eff(phot)}}$ and $T_{\rm{eff(spec)}}$ are both strongly peaked at lower values, but $\Delta T_{\rm{eff(vel)}}$ shows a much wider distribution, and tends towards much larger amplitudes. The amplitudes of $T_{\rm{eff(phot)}}$ and $T_{\rm{eff(spec)}}$ mostly lie between 20 and $200 K$, but $T_{\rm{eff(vel)}}$ amplitudes tend to lie between 200 and $800 K$. The median amplitudes of the sample are 42.3, 72.56, and $582.78 K$ for $\Delta T_{\rm{eff(phot)}}$, $\Delta T_{\rm{eff(spec)}}$ and $\Delta T_{\rm{eff(vel)}}$ respectively. $T_{\rm{eff(vel)}}$, like $T_{\rm{eff(phot)}}$, is a predicted temperature, whereas $T_{\rm{eff(spec)}}$ is measured from our data. $T_{\rm{eff(vel)}}$ has very large amplitudes in the context of radial pulsation. Adding this to the fact that $T_{\rm{eff(vel)}}$ does not agree with $T_{\rm{eff(spec)}}$ suggests that radial pulsation can not explain the LSPs. 

\begin{figure}
\begin{center}
\includegraphics[width=0.5\textwidth]{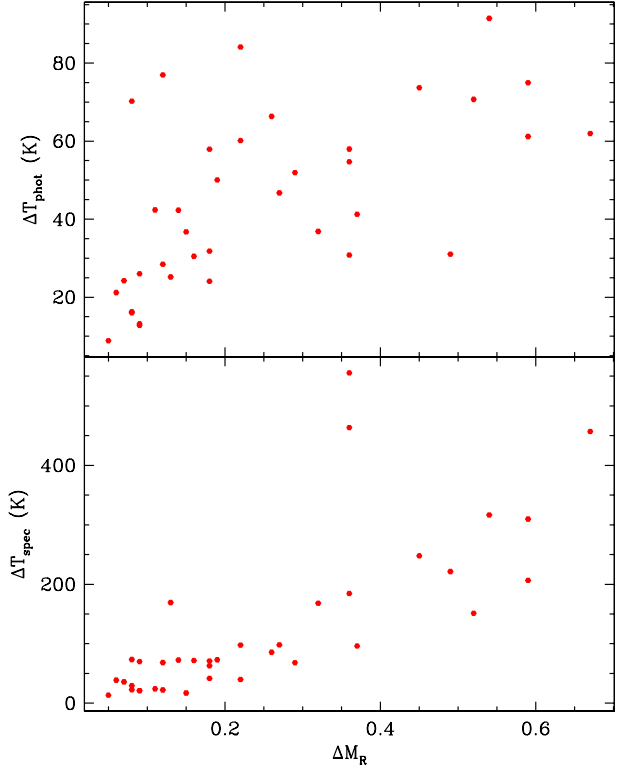}
\end{center}
\caption[Effective temperature amplitude plotted against light amplitude.]{\emph{Top Panel:} $\Delta T_{\rm{eff(phot)}}$ plotted against $\Delta M_R$. \emph{Bottom Panel:} $\Delta T_{\rm{eff(spec)}}$ plotted against $\Delta M_R$.}  
\label{lightamp_tamp}
\end{figure}

Fig.~\ref{lightamp_tamp} shows the relation of light amplitude to $T_{\rm{eff}}$ amplitude for both $T_{\rm{eff(phot)}}$ and
$T_{\rm{eff(spec)}}$. The graph shows that $T_{\rm{eff}}$ amplitude increases with increasing $M_R$ amplitude, for both $T_{\rm{eff(phot)}}$ and
$T_{\rm{eff(spec)}}$.

\section{Discussion and Conclusions}

\subsection{The Movement of the Visible Surface of a Sequence D Star
  during its LSP}

Possibly the most remarkable feature of the sequence D stars is the large change in the radial distance to the visible surface of a sequence
D star as it passes through one LSP\@. By integrating the radial velocity with time we calculated in section~\ref{radial} a radius change for radial pulsation of between 3 and $60 \rm{R_{\odot}}$, with the typical radius full amplitude being $\sim$41 $\rm{R_{\odot}}$. The typical median radius of a sequence D star is around $135 \rm{R_{\odot}}$.

We can use this property to demonstrate just how unlikely radial pulsation is in these stars. An amplitude of $41 \rm{R_{\odot}}$ in a $135 \rm{R_{\odot}}$ radially pulsating star corresponds to a fractional radius change of over 30 per cent from minimum to maximum radius. These radial amplitudes are very large in the context of radial pulsation: fundamental-mode Mira variables pulsate with radius changes of the order of $50R_{\odot}$ \citep{irelandscholzwood}, but they have visible light amplitudes of $\sim$6 magnitudes, while typical sequence D stars have light amplitudes of $\Delta M_R\leq 0.8$ mag. It seems unlikely that such a modest light change could be associated with so large a radius change, suggesting that radial pulsation is not the cause of LSPs.

This comparative analysis backs up a huge problem with the radial pulsation model which was raised in section~\ref{radial}: the changes in radius would lead to changes in $T_{\rm{eff}}$ that are vastly greater than the directly observed changes from spectra or photometric colour.

In addition to the problems with radial pulsation raised by our own data, there are also other problems raised by \cite{sequenceDstars}: the length of the primary period does not vary during the LSP as expected; and to date there is no known mode of radial pulsation that has the required periods.  

\cite{sequenceDstars} proposed nonradial pulsation as a possible cause for the LSPs. They showed that the best match to the observed periods is given by the $g$ mode with $n=2$ and $l=2$. In this case, the nett apparent radial velocity arises from some parts of the visible surface approaching while other parts are receding. Thus at a given position on the stellar surface, the velocity amplitudes and overall radial motions will need to be even larger than in the radial pulsation case. Given the large amplitudes calculated in this work for radial pulsation, nonradial pulsation would lead to distortions of the star that appear far too large to be credible, especially since $g$ modes have significant amplitudes only in radiative regions \citep{sequenceDstars}.

It would therefore seem that the only way to generate the observed change in radial distance to the visible surface of a sequence D star is
to have the star itself move, i.e.~it must be in a binary system. Binary models, however, are inconsistent with various observational results described above. Most particularly, the non-uniform distribution of the angle of periastron and the low companion masses of these stars appear incompatible with a binary system model. The problem of the companion mass might be overcome if the companions were initially planets in eccentric orbits whose mass has been increased via accretion of material from the red giant. This would mean that although the companions currently have masses in the Brown Dwarf range, they were not born as Brown Dwarfs. To remove the problem of the non-uniform distribution of angle of periastron we would need to have an observational selection effect that rejects angles of periastron in the range $0^{\circ}$--$180^{\circ}$, i.e.~we would need an effect that masks the sequence D variation when periastron occurs with the smaller companion in front of the red giant. An effect which could produce this bias is difficult to imagine.

\subsection{Attributes of the Long Secondary Period}

As with past studies, we are unable to find definitive evidence for any model which explains the Long Secondary Periods in sequence D stars. To conclude, we therefore list all the currently known properties of LSPs. Any proposed model for LSPs should be able to explain all of these attributes. 
\begin{enumerate}
\item Stars exhibiting LSPs occupy a clearly defined period-luminosity sequence. 
\item LSPs are of length $\sim$250--1400 days.
\item LSP light variation is not regular and minima in particular vary in depth from cycle to cycle (see Fig. 2 of Wood et al.\ 1999).
\item The primary pulsation is visible in the light curve at all times throughout the LSP and the primary period does not significantly change with LSP phase \citep{sequenceDstars}. 
\item The ratio of LSP to pulsation period is~$\sim$8--10. The shorter period variation lies usually on sequence B and is thought to be the first or second overtone radial pulsation \citep[Wood et al.\ 1999;][]{ita04, fraser05}.
\item Lack of spectroscopic line broadening in observations of sequence D variables indicates that any rotational velocities are $\leq3 \rm{km\,s^{-1}}$ \citep{olivierwood03}. 
\item The radial velocity curves show a characteristic shape: observed radial velocity increases quicker than it decreases.
\item The velocity amplitudes cluster tightly around $3.5 \rm{km\,s^{-1}}$.
\item The light--velocity phase shift for both the short-period and the LSP is $\sim$0.25, with the light lagging the velocity. Minimum light is roughly aligned with mean rising velocity.
\item The ratio of colour to light variation of the LSP and primary oscillations are similar \citep{sequenceDstars}. Similarly, \cite{derekas06} showed that the ratio of blue to red amplitude of the LSP was similar to the ratio for stellar pulsation, and somewhat different from that due to ellipsoidal
light variations.
\item There is no correlation between LSP light and velocity amplitude.
\item Stellar radius tends to be between 100 and $200 \rm{R_{\odot}}$, with large radius variations of 3--$60 \rm{R_{\odot}}$ (if it is assumed that the LSP is caused by radial pulsation).
\item The equivalent width of the H$\alpha$ absorption line varies with the LSP, indicating chromospheric activity \citep{sequenceDstars}.
\end{enumerate}

\section*{Acknowledgments}

PRW has been partially supported in this work by the Australian Research
Council's Discovery Projects funding scheme (project number DP0663447).
We are grateful for the multiple allocations of VLT service time over
several semesters for this extended series of observations (program
identifiers 072.D-0387, 074.D-0098, 075.D-0090 and 076.D-0162).  This
paper utilizes public domain data obtained by the MACHO Project, jointly
funded by the US Department of Energy through the University of
California, Lawrence Livermore National Laboratory under contract
No. W-7405-Eng-48, by the National Science Foundation through the Center
for Particle Astrophysics of the University of California under
cooperative agreement AST-8809616, and by the Mount Stromlo and Siding
Spring Observatory, part of the Australian National University.

\bibliographystyle{mn2e}
\bibliography{bibliographynew}

\begin{thebibliography}{}

\bibitem[\protect\citeauthoryear{{Adams}, {Wood} \& {Cioni}}{{Adams}
  et~al.}{2006}]{betsy}
{Adams} E.,  {Wood} P.~R.,    {Cioni} M.-R.,  2006, Memorie della Societa
  Astronomica Italiana, 77, 537

\bibitem[\protect\citeauthoryear{{Cardelli}, {Clayton} \& {Mathis}}{{Cardelli}
  et~al.}{1989}]{cardelli}
{Cardelli} J.~A.,  {Clayton} G.~C.,    {Mathis} J.~S.,  1989, \apj, 345, 245

\bibitem[\protect\citeauthoryear{{Derekas}, {Kiss}, {Bedding}, {Kjeldsen},
  {Lah} \& {Szab{\'o}}}{{Derekas} et~al.}{2006}]{derekas06}
{Derekas} A.,  {Kiss} L.~L.,  {Bedding} T.~R.,  {Kjeldsen} H.,  {Lah} P.,
  {Szab{\'o}} G.~M.,  2006, \apjl, 650, L55

\bibitem[\protect\citeauthoryear{{Fluks}, {Plez}, {The}, {de Winter},
  {Westerlund} \& {Steenman}}{{Fluks} et~al.}{1994}]{fluks}
{Fluks} M.~A.,  {Plez} B.,  {The} P.~S.,  {de Winter} D.,  {Westerlund} B.~E.,
    {Steenman} H.~C.,  1994, \aaps, 105, 311

\bibitem[\protect\citeauthoryear{{Fraser}, {Hawley} \& {Cook}}{{Fraser}
  et~al.}{2008}]{fraser08}
{Fraser} O.~J.,  {Hawley} S.~L.,    {Cook} K.~H.,  2008, \aj, 136, 1242

\bibitem[\protect\citeauthoryear{{Fraser}, {Hawley}, {Cook} \&
  {Keller}}{{Fraser} et~al.}{2005}]{fraser05}
{Fraser} O.~J.,  {Hawley} S.~L.,  {Cook} K.~H.,    {Keller} S.~C.,  2005, \aj,
  129, 768

\bibitem[\protect\citeauthoryear{{Grether} \& {Lineweaver}}{{Grether} \&
  {Lineweaver}}{2006}]{browndwarfdesert}
{Grether} D.,  {Lineweaver} C.~H.,  2006, \apj, 640, 1051

\bibitem[\protect\citeauthoryear{{Hinkle}, {Fekel}, {Joyce} \& {Wood}}{{Hinkle}
  et~al.}{2006}]{symbioticmira}
{Hinkle} K.,  {Fekel} F.,  {Joyce} R.,    {Wood} P.,  2006, Memorie della
  Societa Astronomica Italiana, 77, 523

\bibitem[\protect\citeauthoryear{{Hinkle}, {Lebzelter}, {Joyce} \&
  {Fekel}}{{Hinkle} et~al.}{2002}]{hinkle02}
{Hinkle} K.~H.,  {Lebzelter} T.,  {Joyce} R.~R.,    {Fekel} F.~C.,  2002, \aj,
  123, 1002

\bibitem[\protect\citeauthoryear{{Hinkle}, {Lebzelter} \& {Scharlach}}{{Hinkle}
  et~al.}{1997}]{hinkle97}
{Hinkle} K.~H.,  {Lebzelter} T.,    {Scharlach} W.~W.~G.,  1997, \aj, 114, 2686

\bibitem[\protect\citeauthoryear{{Hinkle}, {Scharlach} \& {Hall}}{{Hinkle}
  et~al.}{1984}]{hinkle84}
{Hinkle} K.~H.,  {Scharlach} W.~W.~G.,    {Hall} D.~N.~B.,  1984, \apjs, 56, 1

\bibitem[\protect\citeauthoryear{{Houdashelt}, {Bell}, {Sweigart} \&
  {Wing}}{{Houdashelt} et~al.}{2000}]{houdashelt}
{Houdashelt} M.~L.,  {Bell} R.~A.,  {Sweigart} A.~V.,    {Wing} R.~F.,  2000,
  \aj, 119, 1424

\bibitem[\protect\citeauthoryear{{Houk}}{{Houk}}{1963}]{houk}
{Houk} N.,  1963, \aj, 68, 253

\bibitem[\protect\citeauthoryear{{Ireland}, {Scholz} \& {Wood}}{{Ireland}
  et~al.}{2004}]{irelandscholzwood}
{Ireland} M.~J.,  {Scholz} M.,    {Wood} P.~R.,  2004, \mnras, 352, 318

\bibitem[\protect\citeauthoryear{{Ita}, {Tanab{\'e}}, {Matsunaga}, {Nakajima},
  {Nagashima}, {Nagayama}, {Kato}, {Kurita}, {Nagata}, {Sato}, {Tamura},
  {Nakaya} \& {Nakada}}{{Ita} et~al.}{2004}]{ita04}
{Ita} Y.,  {Tanab{\'e}} T.,  {Matsunaga} N.,  {Nakajima} Y.,  {Nagashima} C.,
  {Nagayama} T.,  {Kato} D.,  {Kurita} M.,  {Nagata} T.,  {Sato} S.,  {Tamura}
  M.,  {Nakaya} H.,    {Nakada} Y.,  2004, \mnras, 347, 720

\bibitem[\protect\citeauthoryear{{Keller} \& {Wood}}{{Keller} \&
  {Wood}}{2006}]{kellerwood}
{Keller} S.~C.,  {Wood} P.~R.,  2006, \apj, 642, 834

\bibitem[\protect\citeauthoryear{{Ku{\v c}inskas}, {Hauschildt}, {Ludwig},
  {Brott}, {Vansevi{\v c}ius}, {Lindegren}, {Tanab{\'e}} \& {Allard}}{{Ku{\v
  c}inskas} et~al.}{2005}]{kucinskas}
{Ku{\v c}inskas} A.,  {Hauschildt} P.~H.,  {Ludwig} H.-G.,  {Brott} I.,
  {Vansevi{\v c}ius} V.,  {Lindegren} L.,  {Tanab{\'e}} T.,    {Allard} F.,
  2005, \aap, 442, 281

\bibitem[\protect\citeauthoryear{{Lebzelter}}{{Lebzelter}}{1999}]{lebzelter99}
{Lebzelter} T.,  1999, \aap, 351, 644

\bibitem[\protect\citeauthoryear{{Lebzelter} \& {Hinkle}}{{Lebzelter} \&
  {Hinkle}}{2002}]{lebzelter}
{Lebzelter} T.,  {Hinkle} K.~H.,  2002, \aap, 393, 563

\bibitem[\protect\citeauthoryear{{Lebzelter}, {Kiss} \& {Hinkle}}{{Lebzelter}
  et~al.}{2000}]{lebzelter00}
{Lebzelter} T.,  {Kiss} L.~L.,    {Hinkle} K.~H.,  2000, \aap, 361, 167

\bibitem[\protect\citeauthoryear{{McCarthy} \& {Zuckerman}}{{McCarthy} \&
  {Zuckerman}}{2004}]{mccarthy04}
{McCarthy} C.,  {Zuckerman} B.,  2004, \aj, 127, 2871

\bibitem[\protect\citeauthoryear{{Nicholls}, {Wood} \& {Cioni}}{{Nicholls}
  et~al.}{2009}]{sequenceEstars}
{Nicholls} C.~P.,  {Wood} P.~R.,    {Cioni} M.-R.,  2009, in preparation

\bibitem[\protect\citeauthoryear{{Olivier} \& {Wood}}{{Olivier} \&
  {Wood}}{2003}]{olivierwood03}
{Olivier} E.~A.,  {Wood} P.~R.,  2003, \apj, 584, 1035

\bibitem[\protect\citeauthoryear{{Pasquini} \& {et al.}}{{Pasquini} \& {et
  al.}}{2002}]{pasquinimn}
{Pasquini} L.,  {et al.} 2002, The Messenger, 110, 1

\bibitem[\protect\citeauthoryear{{Payne-Gaposchkin}}{{Payne-Gaposchkin}}{1954}%
]{payne-gaposchkin}
{Payne-Gaposchkin} C.,  1954, Annals of Harvard College Observatory, 113, 189

\bibitem[\protect\citeauthoryear{{Percy}, {Bakos}, {Besla}, {Hou}, {Velocci} \&
  {Henry}}{{Percy} et~al.}{2004}]{percy}
{Percy} J.~R.,  {Bakos} A.~G.,  {Besla} G.,  {Hou} D.,  {Velocci} V.,
  {Henry} G.~W.,  2004, in {Kurtz} D.~W.,  {Pollard} K.~R.,  eds, ASP Conf.
  Ser. 310: IAU Colloq. 193: Variable Stars in the Local Group
  {Multiperiodicity in pulsating red giants}.
pp 348--+

\bibitem[\protect\citeauthoryear{{Press}, {Flannery}, {Teukolsky} \&
  {Vetterling}}{{Press} et~al.}{1986}]{numerical}
{Press} W.~H.,  {Flannery} B.~P.,  {Teukolsky} S.~A.,    {Vetterling} W.~T.,
  1986, Numerical Recipes: The Art of Scientific Computing, 1st edn.
Cambridge University Press, Cambridge (UK) and New York

\bibitem[\protect\citeauthoryear{{Ridgway}, {Joyce}, {White} \&
  {Wing}}{{Ridgway} et~al.}{1980}]{ridgway}
{Ridgway} S.~T.,  {Joyce} R.~R.,  {White} N.~M.,    {Wing} R.~F.,  1980, \apj,
  235, 126

\bibitem[\protect\citeauthoryear{{Scholz} \& {Wood}}{{Scholz} \&
  {Wood}}{2000}]{scholzwood}
{Scholz} M.,  {Wood} P.~R.,  2000, \aap, 362, 1065

\bibitem[\protect\citeauthoryear{{Soszy{\'n}ski}}{{Soszy{\'n}ski}}{2007}]{sosz%
ynski07}
{Soszy{\'n}ski} I.,  2007, \apj, 660, 1486

\bibitem[\protect\citeauthoryear{{Soszynski}, {Dziembowski}, {Udalski},
  {Kubiak}, {Szymanski}, {Pietrzynski}, {Wyrzykowski}, {Szewczyk} \&
  {Ulaczyk}}{{Soszynski} et~al.}{2007}]{oglep-l}
{Soszynski} I.,  {Dziembowski} W.~A.,  {Udalski} A.,  {Kubiak} M.,  {Szymanski}
  M.~K.,  {Pietrzynski} G.,  {Wyrzykowski} L.,  {Szewczyk} O.,    {Ulaczyk} K.,
   2007, Acta Astronomica, 57, 201

\bibitem[\protect\citeauthoryear{{Soszy{\'n}ski} \& {et al. }}{{Soszy{\'n}ski}
  \& {et al. }}{2004}]{ogleellipsoidalmn}
{Soszy{\'n}ski} I.,  {et al. } 2004, Acta Astronomica, 54, 347

\bibitem[\protect\citeauthoryear{{Soszy{\'n}ski}, {Udalski}, {Kubiak},
  {Szymanski}, {Pietrzynski}, {Zebrun}, {Szewczyk} \&
  {Wyrzykowski}}{{Soszy{\'n}ski} et~al.}{2004}]{ogle04}
{Soszy{\'n}ski} I.,  {Udalski} A.,  {Kubiak} M.,  {Szymanski} M.,
  {Pietrzynski} G.,  {Zebrun} K.,  {Szewczyk} O.,    {Wyrzykowski} L.,  2004,
  Acta Astronomica, 54, 129

\bibitem[\protect\citeauthoryear{{Tettelbach} \& {Holdaway}}{{Tettelbach} \&
  {Holdaway}}{2004}]{almanac}
{Tettelbach} F.~M.,  {Holdaway} R.,  2004, The Astronomical Almanac for the
  year 2006.
The Stationery Office, London

\bibitem[\protect\citeauthoryear{{Wood} \& {et al. (MACHO
  Collaboration)}}{{Wood} \& {et al. (MACHO Collaboration)}}{1999}]{wood99mn}
{Wood} P.~R.,  {et al. (MACHO Collaboration)} 1999, in {Le Bertre} T.,  {Lebre}
  A.,   {Waelkens} C.,  eds, IAU Symp. 191: Asymptotic Giant Branch Stars
  {MACHO observations of LMC red giants: Mira and semi-regular pulsators, and
  contact and semi-detached binaries}.
pp 151--+

\bibitem[\protect\citeauthoryear{{Wood}, {Olivier} \& {Kawaler}}{{Wood}
  et~al.}{2004}]{sequenceDstars}
{Wood} P.~R.,  {Olivier} E.~A.,    {Kawaler} S.~D.,  2004, \apj, 604, 800

\end{thebibliography}

\appendix
\section{Calculating Luminosity and Effective Temperature from Photometry}

  The luminosity was calculated from the MACHO Red photometry. The MACHO Red and Blue magnitudes were first dereddened using a colour excess of
  $E(B-V)=0.08$ for the LMC \citep{kellerwood} and absorption $A(\lambda)$ calculated by the reddening law from \cite{cardelli}. This gave
  $A(M_B)=0.2531$ and $A(M_R)=0.1783$. 
  
  The bolometric correction to $M_{R_0}$ (where the zero subscript denotes dereddened values) was derived through the following
  process. Note that we excluded the carbon stars in this procedure since the standard relations do not apply to them. First the
  bolometric correction to $V_0$ was calculated, using a calibration derived from the data of \cite{fluks}: 
  \begin{equation}
    BC_V=0.662-0.872(V-I)_0-0.128(V-I)_0^2,
    \label{BCV}
  \end{equation}
  for $1<(V-I)_0<4$. The colour was converted from $(M_{B}-M_{R})_0$ to
  $(V-I)_0$ using the equation 
  \[
    (V-I)_0=\frac{M_{B{_0}}-M_{R{_0}}-0.120}{0.625}
  \]
  which was derived from MACHO magnitudes and $V$ and $I$ images of our field taken with the 40-inch telescope at Siding Spring Observatory,
  Australia. Substituting $(V-I)_0$ in equation~\ref{BCV} gave $BC_V$.

  The bolometric correction to $M_{B_0}$ was calculated from
  \[
    BC_{M_B}=BC_V-(M_B-V)_0
  \]
  where $V_0$ was calculated using the formula
  \[
    V_0=M_{B_0}+0.042-0.155(M_B-M_R)_0
  \]
  derived again from the 40-inch images. From $BC_{M_B}$ it was a simple
  step to calculate $BC_{M_R}$ using
  \[
    BC_{M_R}=BC_{M_B}-(M_R-M_B)_0,
  \]
  the apparent bolometric magnitude was calculated from
  \[
    m_{\rm{bol}}=M_{R_0}+BC_{M_R},
  \]
  and the absolute bolometric magnitude was calculated from
  \[
    M_{\rm{bol}}=m_{\rm{bol}}-18.54
  \]
  taking the distance modulus as 18.54 for the LMC \citep[from][]{kellerwood}.

  Finally, the luminosity was calculated from
  \[
    L=L_{\odot}\times 10^{0.4(4.75-M_{\rm{bol}})}
  \]
  where 4.75 is the adopted absolute bolometric magnitude of the Sun.

  The effective temperature was calculated from the photometry using a fit made to the stellar atmosphere derived data of \cite{kucinskas} for
  $[Fe/H]=-0.5$ (as appropriate for the LMC):
  \[
    T_{\rm{eff}}=\frac{3723}{((V-I)_0-0.8957)^{\frac{1}{7}}}\qquad\textrm{for }(V-I)_0>1.2.
  \]

\label{lastpage}

\end{document}